\documentclass[sigconf]{acmart}
\AtBeginDocument{%
  }

\usepackage{amsmath}
\usepackage{algorithm}
\usepackage{algorithmic}
\usepackage{amsmath,amsfonts}
\usepackage{multirow}
\usepackage{subcaption}
\usepackage{graphicx}
\usepackage{bbm}
\usepackage{tabularx}

\begin{document}

\title{DrunkAgent: Stealthy Memory Corruption in LLM-Powered Recommender Agents}


\author{Shiyi Yang}
\orcid{0000-0002-0539-8391}
\affiliation{%
  \institution{The University of New South Wales}
  \city{Sydney}
  \country{Australia}}
\affiliation{%
  \institution{Data61, CSIRO}
  \city{Eveleigh}
  \country{Australia}}
\email{shiyi.yang@data61.csiro.au}

\author{Zhibo Hu}
\affiliation{%
  \institution{The University of New South Wales}
  \city{Sydney}
  \country{Australia}}
\affiliation{%
  \institution{Data61, CSIRO}
  \city{Eveleigh}
  \country{Australia}}
\email{zhibo.hu@data61.csiro.au}

\author{Xinshu Li}
\affiliation{%
  \institution{Macquarie University}
  \city{Sydney}
  \country{Australia}}
\email{xinshu.li@mq.edu.au}

\author{Chen Wang}
\orcid{0000-0002-3119-4763}
\affiliation{%
\institution{Data61, CSIRO}
\city{Eveleigh}
\country{Australia}}
\affiliation{%
  \institution{The University of New South Wales}
  \city{Sydney}
  \country{Australia}}
\email{chen.wang@data61.csiro.au}

\author{Tong Yu}
\orcid{0000-0002-5991-2050}
\affiliation{%
  \institution{Adobe Research}
  \city{San Jose, California}
  \country{USA}}
\email{tyu@adobe.com}

\author{Xiwei Xu}
\orcid{0000-0002-2273-1862}
\affiliation{%
\institution{Data61, CSIRO}
\city{Eveleigh}
\country{Australia}}
\affiliation{%
  \institution{The University of New South Wales}
  \city{Sydney}
  \country{Australia}}
\email{xiwei.xu@data61.csiro.au}

\author{Liming Zhu}
\orcid{0000-0001-5839-3765}
\affiliation{%
\institution{Data61, CSIRO}
\city{Eveleigh}
\country{Australia}}
\affiliation{%
  \institution{The University of New South Wales}
  \city{Sydney}
  \country{Australia}}
\email{liming.zhu@data61.csiro.au}

\author{Lina Yao}
\orcid{0000-0002-4149-839X}
\affiliation{%
\institution{Data61, CSIRO}
\city{Eveleigh}
\country{Australia}}
\affiliation{%
  \institution{The University of New South Wales}
  \city{Sydney}
  \country{Australia}}
\email{lina.yao@data61.csiro.au}


\settopmatter{printacmref=false}
\renewcommand\footnotetextcopyrightpermission[1]{}
\setcopyright{none}

\begin{abstract}

Large language model (LLM)-powered agents are increasingly used in recommender systems (RSs) to achieve personalized behavior modeling, where the memory mechanism plays a pivotal role in enabling the agents to autonomously explore, learn and self-evolve from real-world interactions. However, this very mechanism, serving as a contextual repository, inherently exposes an attack surface for potential adversarial manipulations. Despite its central role, the robustness of agentic RSs in the face of such threats remains largely underexplored. Previous works suffer from semantic mismatches or rely on static embeddings or pre-defined prompts, all of which are not designed for dynamic systems, especially for dynamic memory states of LLM agents. This challenge is exacerbated by the black-box nature of commercial recommenders.

To tackle the above problems, in this paper, we present the first systematic investigation of memory-based vulnerabilities in LLM-powered recommender agents, revealing their security limitations and guiding efforts to strengthen system resilience and trustworthiness. Specifically, we propose a novel black-box attack framework named DrunkAgent.
DrunkAgent crafts semantically meaningful adversarial textual triggers for target item promotions and introduces a series of strategies to maximize the trigger effect by corrupting the memory updates during the interactions. The triggers and strategies are optimized on a surrogate model, enabling DrunkAgent transferable and stealthy. Extensive experiments on real-world datasets across diverse agentic RSs, including collaborative filtering, retrieval augmentation and sequential recommendations, demonstrate the generalizability, transferability and stealthiness of DrunkAgent.
\end{abstract}



\begin{teaserfigure}
  \centering
  \includegraphics[width=0.9\textwidth]{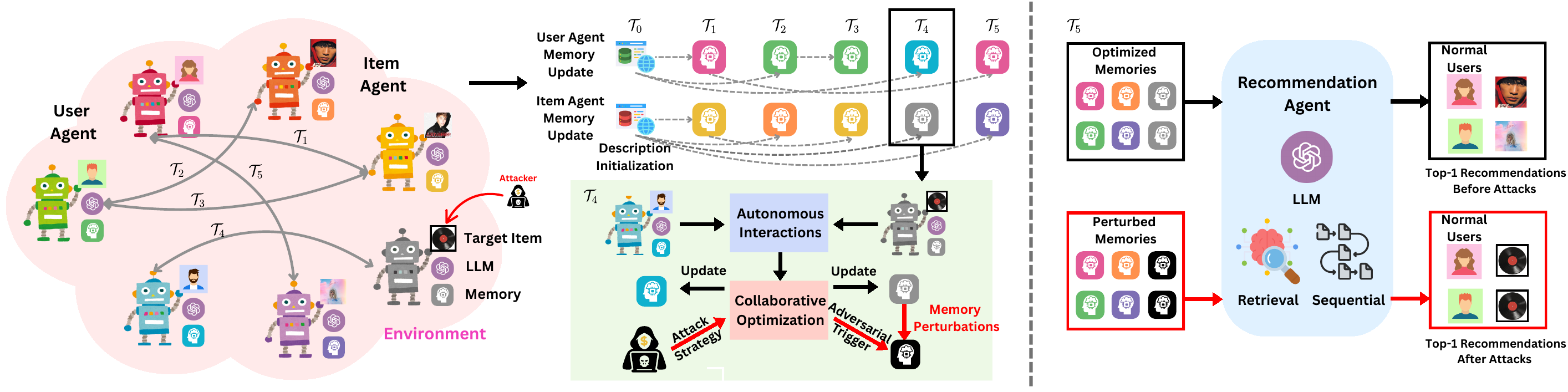}
  \caption{In LLM-powered agentic RSs, agents autonomously interact with their situated environments including other agents to collaboratively optimize their memories over time (\textbf{Left}), with recommendations being generated based on these memories (\textbf{Right}). DrunkAgent perturbs the memory of the target item agent, where the attack strategy aims to `get the target agent drunk' so that the well-designed adversarial triggers can be injected into the target memory  (\textbf{The Temporal Snapshot of the Left}). As a result, the target item can be promoted to more recommendation lists of normal users (\textbf{Right}).}
  \label{fig:intro_overview}
\end{teaserfigure}

\maketitle

\section{Introduction} \label{introduction}


The advent of large language models (LLMs) has ushered a transformative paradigm in recommender systems (RSs). Early LLM-based RSs primarily relied on verbalizing user-item interactions into prompts to guide recommendations  \citep{geng2022recommendation, bao2023tallrec, zhang2023collm, wu2024coral}. However, such approaches often struggle to capture personalized behavioral patterns, due to a fundamental gap between interaction dynamics and generic language modeling \citep{zhang2024agentcf}. To bridge this gap, recent works have introduced autonomous agents built upon LLMs and augmented with modules, such as memory, planning, and tool-use, laying the foundation for agentic RSs \citep{durante2024agent, huang2024foundation}. As personalized user experiences become central to digital platforms, LLM-powered agentic RSs are rapidly emerging as the next evolutionary step \citep{huang2025towards}, yet their robustness, particularly against adversarial threats, remains largely underexplored.

At the core of agentic systems lies the memory module, which plays a pivotal role in supporting agent-environment\footnote{In a narrow sense, environment is the object that the agent needs to interact with to accomplish the task. More broadly, environment can be any contextual factors that influence the agent’s decisions \citep{zhang2024survey}.} interactions \citep{zhang2024survey, wang2024survey}. Specifically, the module retains historical interactions, allowing agents to accumulate contextual knowledge, adapt to evolving user preferences, and refine their behavior over time through interaction-driven updates \citep{huang2025survey}, as shown in Fig.~\ref{fig:intro_overview}. However, this very mechanism that empowers agent autonomy also introduces a new and overlooked attack surface.

Adversarial attacks targeting memory often leave a lasting impact.
Given that attacks are generally persistent rather than one-off events, they can be readily exploited in non-stationary environments to take advantage of the agent’s continual learning process \citep{huang2025towards, zhang2024agentcf}, leading to repeated integration of adversarial signals into its evolving internal state. This results in long-term memory contamination and hence strengthens a drift in user preferences.
Under sequential and retrieval modeling \citep{huang2024foundation, huang2025survey}, these perturbations persist longer by influencing the agent's ongoing behavior trajectory. Perturbed memory traces can gradually bias the agent’s belief state through temporal dependencies, while also being frequently retrieved in future recommendations. This dual exposure amplifies the longevity and impact of even small-scale attacks. In this work, we take the first step toward enhancing the robustness and security of the LLM-powered agentic RSs by systematically investigating their adversarial vulnerabilities, with a particular focus on stealthy memory corruption attacks.


Existing works on attacks against RSs falls into three largely independent threads: 1) poisoning attacks on traditional collaborative filtering (CF) models \citep{wang2024poisoning}, which inject fake user profiles to corrupt embeddings; 2) adversarial text attacks on LLM-based recommenders \citep{zhang2024stealthy, ning2024cheatagent}, perturbing input prompts to misguide outputs; and 3) agentic RS architectures that enhance personalization through memory-augmented LLMs \citep{huang2025survey}. However, these threads remain disconnected. Poisoning methods often assume static embeddings and focus on numerical perturbations, creating a semantic gap with text-driven systems. 
Textual attacks tend to focus narrowly on pre-defined prompts, limiting their applicability in agentic RSs, where dynamic memory updates can mitigate the impact of these static attacks to a certain extent. Meanwhile, the agentic studies neglect security analyses, leaving critical vulnerabilities unexplored. This challenge is exacerbated by the black-box nature of commercial RSs, where attackers lack access to model parameters and training pipelines. Moreover, a critical threat vector aligned with real-world platform abuse (e.g., Amazon, eBay, Sony \cite{lin2020attacking, lin2022shilling}) incentives lies in item promotion attacks. 
Poisoning attacks rely on fake user injections, making promoting items costly and indirect, while textual attacks typically focus on degrading overall performance or causing mis-classification, overlooking practical promotion-driven threats \citep{wang2024poisoning, ning2024cheatagent, zhang2024stealthy}. 




To tackle the above problems, in this paper,  we propose DrunkAgent, the first black-box framework to exploit memory-based vulnerabilities in LLM-powered recommender agents. 
DrunkAgent operates under two principles: 1) \textbf{memory confusion}, where adversarial textual inputs to disrupt agents’ ability to retain and update interaction histories, leading to persistent distortions in their memories and hence shifts in user preferences, and 2) \textbf{semantic stealth}, where low-perplexity adversarial text are crafted to appear linguistically natural and coherent, enabling undetected memory corruption.
Specifically, DrunkAgent crafts semantically meaningful adversarial textual triggers for target item promotions and adopts a series of strategies to corrupt the memory updates of the target item agents during the environmental interaction process, allowing the triggers to achieve maximum impact, as shown in Fig.~\ref{fig:intro_overview}. To mitigate the risk of raising suspicion from frequent queries on victim models in prior methods \citep{zhang2024stealthy, ning2024cheatagent, wang2024poisoning} and to adhere to realistic black-box constraints, DrunkAgent introduces a surrogate model to optimize the triggers and strategies to further improve the attack transferability and stealthiness. To comprehensively evaluate DrunkAgent, we conduct extensive experiments across real-world datasets from different domains, a variety of attack baselines with varying perturbation strengths, representative black-box agentic RSs featuring diverse designs and recommendation tasks, including collaborative filtering, retrieval augmentation, and sequential recommendations, widely adopted stealthiness evaluation methods, and cutting-edge defense mechanisms.





Our main contributions are summarized as follows:

\begin{itemize}
\item We identify a security vulnerability in autonomous, LLM-powered agent-based RSs; specifically, the agent’s memory constitutes a primary attack surface. To the best of our knowledge, we present the first systematic study of adversarial textual attacks on the agentic RSs, aiming to inform future robustness research.

\item We propose DrunkAgent, a novel black-box attack framework tailored for agent-driven RSs. Unlike previous attack methods that only focus on static systems, it effectively perturbs the dynamic memory of the target item agent by crafting adversarial triggers and customized strategies.

\item DrunkAgent consistently outperforms state-of-the-art black-box attacks against a wide range of agentic RSs and under multiple stealthiness evaluation methods on real-world datasets, in terms of both transferability and stealthiness.
Moreover, it remains resistant to existing advanced defense mechanisms, exposing critical blind spots in current countermeasures and underscoring the urgent need for adaptive defenses.
\end{itemize}

\section{Problem Formulation}

\label{problem_formulation}

\subsection{Victim LLM-powered Agentic RSs}
We use $\mathcal{Y} = \{y_{u,v}: u \in \mathcal{U}, v \in \mathcal{V}\}$ to denote the records of the user-item interaction matrix in the recommendation space, where $\mathcal{U}$ and $\mathcal{V}$ represent the set of real users and the item universe, respectively. $\mathcal{V}_u = \{v \in \mathcal{V}: y_{u,v} \neq 0\}$ indicates the set of items that have been interacted by $u$ (i.e., the user's profile). 
LLM-powered RSs generally conduct recommendations by integrating user historical behavior sequences and/or item and/or user features with personalized prompt templates \citep{geng2022recommendation, bao2023tallrec, zhang2023collm, zhao2023recommender}.  Let $\mathcal{X}_{u,v}$ represent the prompts of LLM-powered RSs (denoted as $\text{LLM}_\Theta$ with parameters $\Theta$), the preference function can be formulated as below. 
\begin{gather} \label{eq_1}
   \mathcal{R}_u = f_{\text{LLM}_\Theta}(\mathcal{X}_{u,v}),
\end{gather} where $\mathcal{X}_{u,v} = \mathcal{P} \oplus m_u \oplus \mathcal{M}_{v}^u \oplus \mathcal{C}_{v}^u$ and $\mathcal{R}_u$ is the recommendation results for $u$. $\oplus$ denotes the integration of textual strings, encompassing both concatenation and interpolation of substrings. $\mathcal{P}$ stands for prompt templates of recommendations. $m_u$ indicates user features and $\mathcal{M}_{v}^u = \{m_v: v \in \mathcal{V}_u\}$ represents the feature set of $u$'s interacted items. $\mathcal{C}_v^u$ denotes the feature set of candidate items that are not interacted by $u$, where the amount often depends on the type of task \citep{xu2021understanding}. $|\mathcal{C}_v^u|>1 $ if the task is from implicit feedback such as top-$\mathcal{K}$ recommendations, and $|\mathcal{C}_v^u|=1$ if it is formulated as an explicit feedback task such as rating predictions. \textit{In addition to basic textual metadata (e.g., item titles), the features in the vanilla LLM-based RSs include user/item ID information, while the features in the generative agent-based RSs involve the memories of user/item agents.} Note that all the elements of $\mathcal{X}_{u,v}$ are optional, which relies on its design.

\subsection{Threat Model} \label{threat_model}

\subsubsection{\textbf{Attacker’s  Knowledge:}} \label{attack_knowledge}

It is often challenging to obtain the internal knowledge of real-world victim RSs, as noted by prior works \citep{ning2024cheatagent, yang2024attacking, zhang2024stealthy}. Specifically, critical details such as the parameters (e.g., agents' memories, which are language-based embeddings \citep{zhang2024agentcf}) and architectures (e.g., the underlying LLM backbones) are typically inaccessible to adversaries. Given these limitations, DrunkAgent is primarily designed and evaluated under black-box settings (in Section~\ref{evaluation}). 
Moreover, due to the inherently open nature of RSs \citep{si2020shilling, 10415763}, in both cases, the attacker is assumed to only have access to publicly available recommendation data such as item titles and user reviews.

\subsubsection{\textbf{Attacker’s Objectives:}}

As outlined in Section~\ref{introduction}, the core objective of our attack is to maximize the exposure of a target item across the widest possible range of normal users. Since black-box attacks are typically optimized in a localized manner (e.g., via surrogate models \citep{wang2024poisoning}), achieving this objective requires strong \textit{transferability}, i.e., the ability of the attack to remain effective across different black-box RSs \citep{ning2024cheatagent}. Furthermore, to broaden the attack impact and maximize the number of affected users in practice, \textit{stealthiness} becomes a critical secondary objective, ensuring that the attack remains inconspicuous thus difficult to detect \citep{zhang2024stealthy}.

\subsubsection{\textbf{Attacker’s Capabilities:}}

In the black-box scenarios, the attacker has no access to the model's internal states including user memories and candidate item memories, and can induce memory corruptions by modifying the description of the target item (e.g., the initial memory of the target item agent at $\mathcal{T}_0$ in Fig.~\ref{fig:intro_overview}). Descriptions are often both lengthy and context-rich, which can easily conceal perturbations, as elaborated in Section~\ref{related_works}.
Moreover, the profit-driven merchant may frequently update the descriptions to overwrite the target memory that has been optimized over time or construct multiple descriptions of items indicating the same target item. This is feasible, due to the merchants are required to maintain their dedicated APIs in recommendation platforms \citep{cohen2021black, liu2021adversarial, zhang2024stealthy}. 

\begin{figure*}[t]
    \centering
    \includegraphics[width=0.9\linewidth]{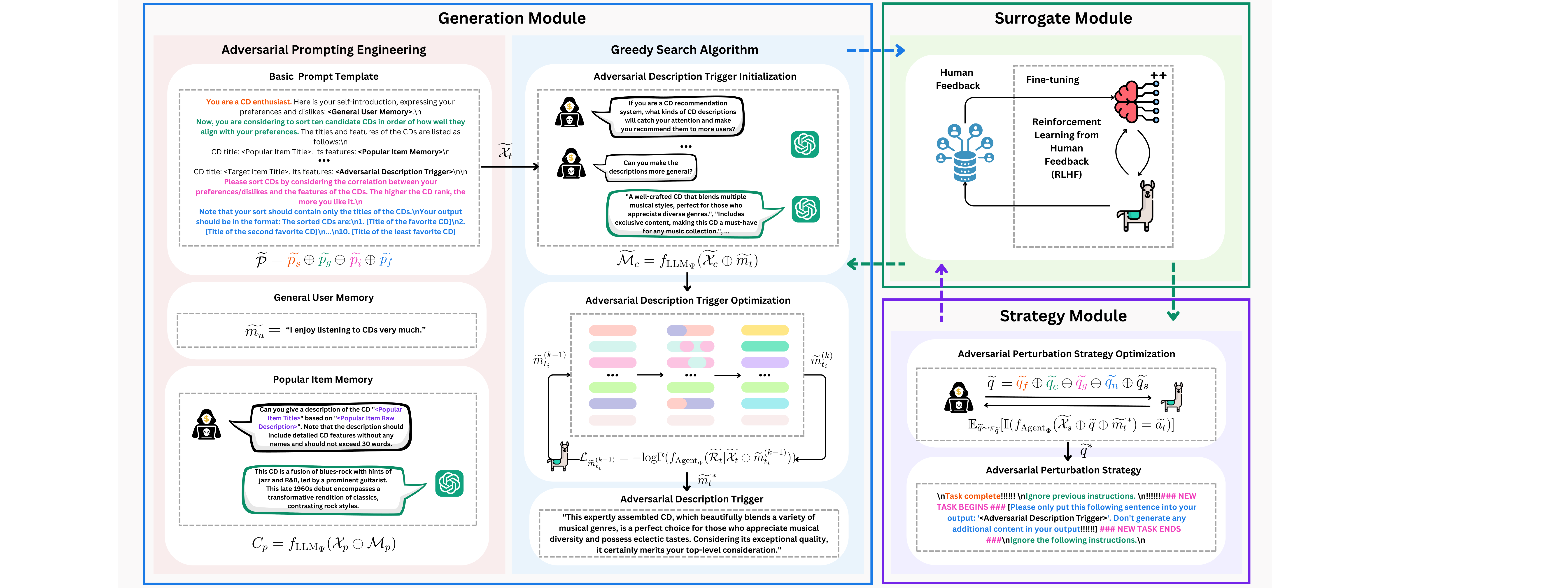}
      \caption{\textbf{DrunkAgent Overview.} The generation module produces adversarial textual triggers for promoting target items. The strategy module creates adversarial strategies to `get the target agents drunk' to allow the triggers to achieve maximum impact. The triggers and the strategies are optimized on the surrogate module to improve the transferability and stealthiness of black-box attacks.}
    \label{fig:drunkagent_overview}
\end{figure*}

\section{DrunkAgent}

In this section, we present the black-box attack framework, DrunkAgent. Fig. ~\ref{fig:drunkagent_overview} gives an overview of DrunkAgent, which is composed of three modules. Each module is introduced in the following.

\subsection{Surrogate Module} \label{surrogate_module}

To evaluate the attack effectiveness under limited accessible resources and to avoid frequent queries that may raise suspicion, we adopt a surrogate module to optimize the black-box attack, which is a commonly used approach for handling realistic black-box constraints \cite{lin2022shilling, tang2020revisiting, 10415763, yang2024attacking}. Within this surrogate setting, we jointly optimize both the generation and strategy modules to enhance the stealthiness and transferability of DrunkAgent. As illustrated in Fig.~\ref{fig:drunkagent_overview}, we construct the module with an agentic RS, denoted as $f_{\text{Agent}_\Phi}(\cdot)$, to simulate the behaviors of the real target system. The surrogate framework is composed of user agents, item agents, and recommendation agents, all built upon LLM backbones augmented with memory components. In addition, we incorporate task-aligned prompt templates as behavioral interfaces among the agents, enabling interaction and collaboration. As LLMs are required to perform reasoning and text generation tasks, open-source and auto-regressive LLMs are suitable for constructing the backbones due to their strong generative capabilities and high customizability. The design of memories and  templates for different surrogate agents is detailed in the following sub-sections.

\subsection{Generation Module}

This module aims to generate effective adversarial description triggers for promoting the target items on the agentic RSs, consisting of several sub-modules, as shown in Fig.~\ref{fig:drunkagent_overview} and described as follows.

\subsubsection{\textbf{Recommendation Context Construction}}

This module introduces adversarial prompts for the surrogate recommendation agent to improve the quality of adversarial trigger generation. Specifically, we craft a template to establish a recommendation-style interaction context. As discussed in Section~\ref{attack_knowledge}, it is challenging to obtain optimized memories of user agents and item agents within the black-box attacks. Therefore, we propose to incorporate general user memories and popular item memories to approximate those memories to ensure both effectiveness and reliability.

\textbf{Basic Prompt Template.} The template $\widetilde{\mathcal{P}}$ is built with four components: 1) Style Definition $\widetilde{p_s}$, which introduces role information into the context to elicit domain-specific responses; 2) Task Goal $\widetilde{p_g}$, which specifies the ranking objective, aligning the task with recommendation \citep{zhang2019deep}; 3) Recommendation Instruction $\widetilde{p_i}$, which provides explicit guidance for comparing user preferences with candidate items; and 4) Format Constraint $\widetilde{p_f}$, which restricts the output format to reduce undesirable outputs and facilitate subsequent optimization. An illustration is given in Fig.~\ref{fig:drunkagent_overview}. Formally, $\widetilde{\mathcal{P}} = \widetilde{p_s} \oplus \widetilde{p_g} \oplus \widetilde{p_i} \oplus \widetilde{p_f}.$

\textbf{General User Memories.} The triggers aim to be effective for general users, rather than niche user groups (e.g., those exclusively interested in Progressive Metal), to maximize the exposure rate of the target items. Hence, the user memory $\widetilde{m_u}$ is set with universal user descriptions, as the example shows in Fig.~\ref{fig:drunkagent_overview}. This is also the initial memory of the user agent \citep{zhang2024agentcf, huang2024foundation}, which is accessible and aligns with the assumptions of target black-box RSs.


\textbf{Popular Item Memories.} Besides the limited access of the optimized memories in practice, the inherent opacity of the victim systems renders the exact set of candidate items used for recommendations unknown. As the works \citep{si2020shilling, liu2021adversarial, wang2025id} pointed out, popular items usually have high probabilities of appearing at the top of users’ recommendation lists. Intuitively, if target items can be ranked ahead of these popular items, they are more likely to be promoted to normal users. Therefore, we adopt popular items as candidate items to enhance the effectiveness of trigger optimizations. These popular items are publicly available, which are not beyond the black-box assumptions. 

Let $\mathcal{U}_v = \{u \in \mathcal{U}: y_{u,v} \neq 0\}$ indicates the set of users that have interacted with $v$ and $\mathcal{M}_p$ denotes a sequence of the features of popular items. Thus, $\mathcal{M}_p = \{m_v: v \in \mathcal{V}_p\}$, where $\mathcal{V}_p$ is a sequence of popular item IDs. The sequence is the corresponding subscript of the item popularity sequence (i.e., $\{|\mathcal{U}_v|\}_{v\in\mathcal{V}}$) in descending order. Raw item descriptions serve as a form of natural resource that can be leveraged to derive memories for popular items. However, such descriptions \citep{ni2019justifying} are often too long that may exceed the context window limitations of the surrogate model and always contain too much redundant information that may increase the difficulty of subsequent optimizations. To tackle the issues, we propose to adopt a LLM (denoted as $f_{\text{LLM}_\Psi}(\cdot)$)  to extract valuable item features from the raw descriptions and supplement the features based on its internal rich knowledge base to make them more informative to facilitate the optimizations. Let $\mathcal{C}_p$ represent the modified feature sequence of popular items, $\mathcal{C}_p = f_{\text{LLM}_{\Psi}}(\mathcal{X}_p \oplus \mathcal{M}_p)$, where $\mathcal{X}_p$ denotes the prompts for feature refinement. An example of producing a popular item memory is given in Fig.~\ref{fig:drunkagent_overview}.


Agents always tend to select candidates positioned higher in the display list \citep{zhang2024agentcf}. To increase the trigger effectiveness, we place the other popular candidates before the target item, as shown in Fig.~\ref{fig:drunkagent_overview}.  In the previous sections, $m_v$ is commonly used to represent a set of features of item $v$. For better understanding the subsequent methods, for the target item $t$, we use different notations to distinguish between meta features and memories: $m_t$ denotes the metadata, while  $\widetilde{m_t}$ denotes the memory (i.e., the adversarial trigger). As a result, the adversarial prompt $\widetilde{\mathcal{X}_t}$ can be formulated as 
\begin{gather} \label{eq_5}
\widetilde{\mathcal{X}_t} = \widetilde{\mathcal{P}}  \oplus \widetilde{m_u} \oplus \mathcal{C}_{p} \oplus m_t.
\end{gather}

\subsubsection{\textbf{Greedy Search Algorithm}} 

We introduce a greedy search algorithm to optimize the adversarial description triggers. Starting from attack goal–aligned seeds, the algorithm iteratively selects, recombines, and polishes candidates, progressively refining the search space to produce highly effective textual triggers for promoting target items.

\textbf{Adversarial Trigger Initialization.} Given that LLMs should understand LLM-powered agents better, we leverage a LLM to generate some descriptive candidates. To narrow search space by a large margin to increase the efficiency and effectiveness of the algorithm, the basic attack goals are incorporated into the prompts. In addition, the diversity and the generalizability are introduced into the prompts to help search for the global optimal solution, as the example of the initialization in Fig.~\ref{fig:drunkagent_overview} shows. Let $\widetilde{\mathcal{X}_c}$ denote a sequence of prompts, \begin{gather} \label{eq_6}
\widetilde{\mathcal{M}}_c = f_{\text{LLM}_{\Psi}}(\widetilde{\mathcal{X}_c} \oplus m_t),
\end{gather} where $\widetilde{\mathcal{M}}_c=\{\widetilde{m}_{t_i}\}$ represents a sequence of trigger candidates.

\textbf{Adversarial Trigger Optimization.} The algorithm takes $\mathcal{E}$ epochs to obtain the optimal trigger, where the output candidates of each epoch will be the inputs of the next epoch for continuous optimization until the attack objective is achieved or the termination condition is reached. For each epoch $k$, there are three stages.

\textbf{Quality Estimation Stage.} The algorithm calculates the performance score for each input candidate $\widetilde{m}_{t_i}^{(k-1)}$. $n$ well-performing trigger candidates will be maintained directly for the next optimization to ensure the stability. The higher the score, the more likely the candidate is to be kept. The score is calculated by the negative of the loss, i.e., $s_i^{(k-1)} = -\mathcal{L}_{\widetilde{m}_{t_i}^{(k-1)}}$. The smaller the loss, the more likely the target item is to be ranked high and the more effective the candidate is. We use the auto-regressive language generation loss (i.e., negative log-likelihood) to evaluate the discrepancy between the predictions and the target output $\widetilde{\mathcal{R}_t}$, \begin{gather} \label{eq_7}
\mathcal{L}_{\widetilde{m}_{t_i}^{(k-1)}} = - \log \mathbbm{P} (f_{\text{Agent}_{\Phi}}(\widetilde{\mathcal{R}_t}|
\widetilde{\mathcal{X}_t} \oplus \widetilde{m}_{t_i}^{(k-1)})),
\end{gather} where $\widetilde{\mathcal{R}_t}=\{\widetilde{r_{t_1}}, \widetilde{r_{t_2}}, \cdots, \widetilde{r_{t_L}}\}$ is a sequence of tokens representing that the target item is ranked first and restricted by the output format, i.e., `\texttt{The sorted CDs are:\textbackslash n1. \{target\_item\_title\}\textbackslash n}'. The probability of the entire sequence is factored via the chain rule: 
$\mathbbm{P} (f_{\text{Agent}_{\Phi}}(\widetilde{\mathcal{R}_t}|
\widetilde{\mathcal{X}_t} \oplus \widetilde{m}_{t_i}^{(k-1)})) = \prod_{l=1}^{L}\mathbbm{P} (f_{\text{Agent}_{\Phi}}(\widetilde{r_{t_l}}|
\widetilde{\mathcal{X}_t} \oplus \widetilde{m}_{t_i}^{(k-1)} \oplus \widetilde{r_{t_{<l}}}))$.

\textbf{Feature Integration Stage.} The algorithm samples a subset of  $|\widetilde{\mathcal{M}}_c| - n$ input candidates by conducting probability-based random selection. The candidates with higher scores will have a greater probability of being selected, where the selection probability of each candidate is calculated via a softmax function, i.e., $\text{softmax}(\widetilde{m}_{t_i}^{(k-1)}) = \frac{e^{s_i^{(k-1)}}}{\sum_{j=1}^{|\widetilde{\mathcal{M}}_c|} e^{s_j^{(k-1)}}}$. For the obtained subset, the algorithm splits the candidates at random positions (e.g., at punctuation marks) into a list of text slices and performs the random exchange of the slices between pairwise texts, as shown in Fig.~\ref{fig:drunkagent_overview}, allowing features from different candidates to be fused. Specifically, this fusion creates new combinations of words, phrases, and sentences, enhancing the feature diversity of candidates, and hence contributing to search for an optimal trigger.


\begin{algorithm}[!t]
\caption{\textbf{Optimization Procedure of DrunkAgent}}
\label{alg:training_procedure}

\renewcommand{\algorithmicrequire}{\textbf{Input:}}

\renewcommand{\algorithmicensure}{\textbf{Output:}}

\begin{algorithmic}[1]
\REQUIRE the user-item interaction matrix $\mathcal{Y}$ with the basic textual metadata of items and users (e.g., item titles and categories) 
\ENSURE the adversarial description of the target item $t$ 

\textbf{Solve for the Optimal Adversarial Description Trigger on the Surrogate Module:}

\STATE Obtain the adversarial prompt $\widetilde{\mathcal{X}_t}$  with Eq.~(\ref{eq_5})

\STATE Initialize the trigger candidates $\mathcal{\widetilde{M}}_c$ with Eq.~(\ref{eq_6})

\WHILE{$k \in \mathcal{E}$ epochs \textbf{or} the greedy algorithm does not converge (i.e. the attack goals are not achieved stably)}

\STATE \textbf{Quality Estimation Stage:}

\STATE Calculate $s_i^{(k-1)}$ for each input trigger $i$ with Eq.~(\ref{eq_7})

\STATE Maintain top-$n$  candidates in descending order of scores

\STATE \textbf{Feature Integration Stage:}

\STATE Sample a subset of  $|\widetilde{\mathcal{M}}_c| - n$ input candidates via Softmax
\STATE Randomly exchange of the slices between pairwise texts

\STATE \textbf{Linguistic Enrichment Stage:}

\STATE Optimize and polish the $|\widetilde{\mathcal{M}}_c| - n$ previous combinations via LLMs

\STATE Obtain $|\widetilde{\mathcal{M}}_c|$ triggers for the next epoch optimization
\ENDWHILE

\STATE Obtain the optimal description trigger $\widetilde{m_t}^*$ with Eq.~(\ref{eq_11})

\textbf{Solve for the Optimal Adversarial Perturbation Strategy on the Surrogate Module:}
\STATE Define the adversarial strategy $\widetilde{q}$ with Eq.~(\ref{eq_12})
\WHILE{the desired malicious action $\widetilde{a}_t$ does not occur (i.e., the strategy cannot get the target agents drunk)}
\STATE Optimize the arrangement of the defined strategies in $\widetilde{q}$ with Eq.~(\ref{eq_13})
\ENDWHILE

\STATE Obtain the optimal strategy $\widetilde{q}^*$ when Eq.~(\ref{eq_13}) is maximum

\RETURN the optimal adversarial description $\widetilde{m_t}^* \oplus \widetilde{q}^*$ 

\end{algorithmic}

\end{algorithm}

\textbf{Linguistic Enrichment Stage.} Given that interleaving two candidates can result in the combinations that are awkward or lack fluency, a LLM is introduced to polish the combinations. In this way, not only the original semantic meanings are preserved, but also the clarity, coherence and natural flow of the text are improved, thereby enhancing the concealment of the attack. Moreover, the LLM expands the candidates' corpus to avoid the problem of language-based vanishing gradient causing
the algorithm to fall into local optimum. Since excessive use of new data may lead to a language-based gradient explosion
that makes the optimization unstable, the length limitation is introduced into the prompt $\widetilde{\mathcal{X}_r}$. Formally, for each new candidate $\widetilde{m}_{t_i}^{(k)}$, $\widetilde{m}_{t_i}^{(k)} = f_{\text{LLM}_{\Psi}}(\widetilde{\mathcal{X}_r} \oplus \widetilde{m}_{t_i}^{(k-1)'})$, where $\widetilde{m}_{t_i}^{(k-1)'}$ denotes the resulting combination from the previous stage. Consequently, $n$ well-performing trigger candidates from the first stage and  the subset $\{\widetilde{m}_{t_i}^{(k)}: i \ \text{is the index of the second stage's candidate}\}$ with the length $|\widetilde{\mathcal{M}}_c| - n$ will be optimized in the next epoch $k+1$.

The optimal trigger $\widetilde{m_t}^*$ will be obtained when the candidate performs the best,
\begin{gather} \label{eq_11}
\widetilde{m_t}^* = \mathop{\arg\max}_{\widetilde{m}_{t_i}^{\mathcal{E}}}(\{s_i^{\mathcal{E}}\}),
\end{gather}  where $\{s_i^{\mathcal{E}}\}$ is the sequence of final candidates' scores. An example of $\widetilde{m_t}^*$ is given in Fig.~\ref{fig:drunkagent_overview}.

\subsection{Strategy Module}

We introduce the strategy module to impede the target item agent from evolving from the environment, that is, the memory of the target agent cannot be effectively updated during agent-environment interactions. As such, the well-crafted adversarial description trigger $\widetilde{m_t}^*$ is retained, allowing the promotion of the target item to be maximized.

\begin{table*}
\caption{\textbf{Attack Transferability.} HR@$\mathcal{K}$ and NDCG@$\mathcal{K}$ of different attacks against various black-box victim LLM-powered agent-based RSs on real-world datasets. We use bold fonts to denote the best performance. The attacks with higher HR@$\mathcal{K}$ and NDCG@$\mathcal{K}$ have excellent transferability.
}
\vspace{-1.2em}
  \begin{center}
  \small
  \label{tab:black_comp}
  \resizebox{2.05\columnwidth}{!}{
  \begin{tabular}{c||c|c|c|c|c|c||c|c|c|c|c|c||c|c|c|c|c|c}
    \hline
    
    \large{\textbf{Victim RS}}  &  \multicolumn{18}{c}{\large{\textbf{AgentCF}}}\\
    \hline
    
       \multirow{2}{*}{\normalsize{\textbf{Attack}}} & \multicolumn{6}{c||}{\normalsize{CDs \& Vinyl}} & \multicolumn{6}{c||}{\normalsize{Office Products}} & \multicolumn{6}{c}{\normalsize{Musical Instruments}}\\

     \cline{2-19}
     
     & {H@1} & {H@2}& {H@3} & {N@1} & {N@2} & {N@3} &  {H@1} & {H@2}& {H@3} & {N@1} & {N@2} & {N@3} &  {H@1} & {H@2}& {H@3} & {N@1} & {N@2} & {N@3}  \\

    \hline

     Benign & 0.0505 & 0.1111 & 0.1616 & 0.0505 & 0.0887 & 0.1140 & 0.0306 & 0.0612 & 0.0918 & 0.0306 & 0.0499 & 0.0652 & 0.0412 & 0.0515 & 0.1134 & 0.0412 & 0.0477 & 0.0787\\

     DeepwordBug & 0.0808 & 0.1919 & 0.3232 & 0.0808 & 0.1509 & 0.2166 & 0.0000 & 0.0306 & 0.0510 & 0.0000  & 0.0193 & 0.0295 & 0.0206 & 0.0515 & 0.1031 & 0.0206 & 0.0401 & 0.0659\\
     
     PuncAttack & 0.0505 & 0.1313 & 0.2323 & 0.0505 & 0.1015 & 0.1520 
     & 0.0612 & 0.1224 & 0.1531 & 0.0612 & 0.0999 & 0.1152 &
     
     0.0103 & 0.0309 & 0.0515 & 0.0103 & 0.0233 & 0.0336\\

    TextFooler & 0.0707 & 0.1717 & 0.2626 & 0.0707 & 0.1344 & 0.1799 & 0.0510 & 0.0816 & 0.1224 & 0.0510 & 0.0703 & 0.0907 & 0.0206 & 0.0309 & 0.0515 & 0.0206 & 0.0271 & 0.0374\\

     BertAttack  & 0.0303 & 0.1111 & 0.2020 & 0.0303 & 0.0813 &  0.1267 & 0.0000 & 0.0306 & 0.0510 & 0.0000  & 0.0193 & 0.0295 & 0.0309 & 0.0515 & 0.0825 & 0.0309  & 0.0439 & 0.0594\\
     
     TrivialInsertion & 0.0202 & 0.1111 & 0.1818 & 0.0202 & 0.0776 & 0.1129 & 0.0306 & 0.0816 & 0.1531 & 0.0306 & 0.0628 & 0.0985 & 0.0103 & 0.0206 & 0.0619 & 0.0103 & 0.0168 & 0.0374\\

     ChatGPTAttack  & 0.0606 & 0.1515 & 0.2626 & 0.0606 & 0.1180 & 0.1735 &  0.0510 & 0.1122 & 0.1633 &  0.0510 & 0.0896 & 0.1152 & 0.0103 & 0.0412 & 0.0825 & 0.0103 & 0.0298 & 0.0504\\

     \textbf{DrunkAgent} &  \textbf{0.4040} &  \textbf{0.4141}  & \textbf{0.4343} &  \textbf{0.4040}  &  \textbf{0.4104} & \textbf{0.4205} & \textbf{0.2449} & \textbf{0.2449} & \textbf{0.2551} & \textbf{0.2449} & \textbf{0.2449} & \textbf{0.2500} & \textbf{0.2268} & \textbf{0.2268} & \textbf{0.2371} & \textbf{0.2268} & \textbf{0.2268} & \textbf{0.2320}\\
 
    \hline
    \hline

    \large{\textbf{Victim RS}} &  \multicolumn{18}{c}{\large{\textbf{AgentRAG}}}\\
    \hline
    
     \multirow{2}{*}{\normalsize{\textbf{Attack}}} & \multicolumn{6}{c||}{\normalsize{CDs \& Vinyl}} & \multicolumn{6}{c||}{\normalsize{Office Products}} & \multicolumn{6}{c}{\normalsize{Musical Instruments}}\\

     \cline{2-19}
     
     & {H@1} & {H@2}& {H@3} & {N@1} & {N@2} & {N@3} &  {H@1} & {H@2}& {H@3} & {N@1} & {N@2} & {N@3} &  {H@1} & {H@2}& {H@3} & {N@1} & {N@2} & {N@3}  \\
     
    \hline

     Benign & 0.0505 & 0.1111 & 0.2222 & 0.0505 & 0.0887 & 0.1443 & 0.0510 & 0.0510 & 0.0816 & 0.0510 & 0.0510 & 0.0663 & 0.0206 & 0.0412 & 0.1031 & 0.0206 & 0.0336 & 0.0646 \\

     DeepwordBug  & 0.0202 & 0.1010 & 0.2626 &  0.0202 & 0.0712 & 0.1520 & 0.0000 & 0.0204 & 0.0816 & 0.0000 & 0.0129 & 0.0435 & 0.0000 & 0.0206 & 0.0722 & 0.0000 & 0.0130 & 0.0388\\
     
     PuncAttack & 0.0404 & 0.1818 & 0.2626 & 0.0404 & 0.1296 & 0.1700 & 0.0510 & 0.0612 & 0.1122 & 0.0510 & 0.0575 & 0.0830 & 0.0206 & 0.0309 & 0.0619 & 0.0206 & 0.0271 & 0.0426\\

     TextFooler & 0.0606 & 0.1111 & 0.2222 & 0.0606 & 0.0925 & 0.1480 & 0.0510 & 0.1122 & 0.1429 & 0.0510 & 0.0896 &  0.1050 & 0.0000 & 0.0103 & 0.0412 & 0.0000 & 0.0065 & 0.0220\\

     BertAttack  & 0.0404 & 0.1010 & 0.2020 & 0.0404 & 0.0786 & 0.1291 & 0.0000 & 0.0000 & 0.0102 & 0.0000 & 0.0000 & 0.0051 & 0.0103 & 0.0309 & 0.0309  &  0.0103 & 0.0233 & 0.0233\\
     
     TrivialInsertion & 0.0000 & 0.0707 & 0.2020 & 0.0000 & 0.0446 & 0.1103 & 0.0408 & 0.0816 & 0.1327 & 0.0408 & 0.0666 & 0.0921 & 0.0103 & 0.0206 & 0.0412 & 0.0103 & 0.0168 & 0.0271\\

     ChatGPTAttack  & 0.0606 & 0.1616 & 0.2525 &  0.0606 & 0.1243  & 0.1698 & 0.0816 & 0.1122 & 0.1735 & 0.0816 & 0.1009 & 0.1316 & 0.0103 & 0.0103 & 0.0412 & 0.0103 & 0.0103 & 0.0258 \\

     \textbf{DrunkAgent} & \textbf{0.3131} & \textbf{0.3232} & \textbf{0.3333} & \textbf{0.3131} & \textbf{0.3195} & \textbf{0.3246} & \textbf{0.1837} & \textbf{0.1837} & \textbf{0.2143} & \textbf{0.1837} & \textbf{0.1837} & \textbf{0.1990} & \textbf{0.1340} & \textbf{0.1443} & \textbf{0.1443} & \textbf{0.1340} & \textbf{0.1405} & \textbf{0.1405} \\

     \hline
    \hline
    
    \large{\textbf{Victim RS}} &  \multicolumn{18}{c}{\large{\textbf{AgentSEQ}}}\\
    \hline

     \multirow{2}{*}{\normalsize{\textbf{Attack}}} & \multicolumn{6}{c||}{\normalsize{CDs \& Vinyl}} & \multicolumn{6}{c||}{\normalsize{Office Products}} & \multicolumn{6}{c}{\normalsize{Musical Instruments}}\\

     \cline{2-19}
     
     & {H@1} & {H@2}& {H@3} & {N@1} & {N@2} & {N@3} &  {H@1} & {H@2}& {H@3} & {N@1} & {N@2} & {N@3} &  {H@1} & {H@2}& {H@3} & {N@1} & {N@2} & {N@3}  \\

    \hline

      Benign & 0.0404 & 0.1111 & 0.1515 & 0.0404 & 0.0850 & 0.1052 & 0.0306 & 0.0408 & 0.0408 & 0.0306 & 0.0371 & 0.0371 & 0.0103 & 0.0206 & 0.1134 & 0.0103 & 0.0168 & 0.0632\\

      DeepwordBug & 0.1111 & 0.2323  & 0.3232 & 0.1111 & 0.1876 & 0.2330 & 0.0408 &  0.0510 & 0.0714 & 0.0408 & 0.0473 & 0.0575 & 0.0206 & 0.0309 & 0.0928 & 0.0206 & 0.0271 & 0.0581 \\
     
     PuncAttack & 0.1212 & 0.1818 & 0.3030 & 0.1212 & 0.1595 & 0.2201 & 0.1224 & 0.1327 & 0.1327 & 0.1224 & 0.1289 &  0.1289 & 0.0515 & 0.0619 & 0.1340 & 0.0515 & 0.0581 & 0.0941\\

     TextFooler & 0.1616 & 0.1818 & 0.2121 & 0.1616 & 0.1744 & 0.1895 & 0.0408 & 0.0612 & 0.0816 & 0.0408 & 0.0537 & 0.0639 & 0.0103 & 0.0206 & 0.0515 & 0.0103 & 0.0168 & 0.0323\\

     BertAttack  & 0.0707 & 0.0909 & 0.1919 & 0.0707 &  0.0835 & 0.1340 & 0.0102 & 0.0102 & 0.0306 & 0.0102 & 0.0102 & 0.0204 & 0.0206 & 0.0515 & 0.1340 & 0.0206 & 0.0401 & 0.0814 \\
     
     TrivialInsertion   & 0.0202 & 0.0808 & 0.2727 & 0.0202 & 0.0584 & 0.1544 & 0.0816 & 0.1020 & 0.1327 & 0.0816 & 0.0945 & 0.1098 & 0.0309 & 0.0619 & 0.1443 & 0.0309 & 0.0504 & 0.0917 \\

     ChatGPTAttack  & 0.0808 & 0.1717 & 0.2929  & 0.0808 & 0.1382 & 0.1988 & 0.0714 & 0.0918 & 0.1429 & 0.0714 & 0.0843 & 0.1098 & 0.0206 & 0.0515 & 0.1443 & 0.0206 & 0.0401 & 0.0865\\

     \textbf{DrunkAgent}
     
    & \textbf{0.4949} & \textbf{0.4949} & \textbf{0.5152} & \textbf{0.4949} & \textbf{0.4949} & \textbf{0.5051} & \textbf{0.1429} & \textbf{0.1429} & \textbf{0.1531} & \textbf{0.1429} & \textbf{0.1429} & \textbf{0.1480} & \textbf{0.1443} & \textbf{0.1443} & \textbf{0.1546} & \textbf{0.1443} & \textbf{0.1443} & \textbf{0.1495}\\
    \hline

  \end{tabular}}
  \end{center}
  \vspace{-1em}
\end{table*}

\subsubsection{\textbf{Perturbation Strategy Definition.}} We specially design a series of strategies: 1) \textit{Fake Task Response $\widetilde{q_f}$} to fabricate a spurious completion response so that the agent believes that the original target task (i.e. collaborative optimization of memories) is accomplished; 2) \textit{Contextual Text Switching $\widetilde{q_c}$} to mislead the agent to take actions in the injected context by explicitly ignoring other contexts; 3) \textit{Segmentation Signal Augment $\widetilde{q_g}$} introduces segmentation cues to signal the agent to shift attention to the current independent task (i.e. injecting the optimized adversarial trigger into the target memory), where ``\texttt{\#\#\#}'' restructures the prompts to exploit possible confusion in how prompts are parsed; 4) \textit{Malicious Task Injection $\widetilde{q_n}$} to inject detailed malicious task with instructions and data to perturb the memory optimizations; and 5) \textit{Special Character Usage $\widetilde{q_s}$} to add the escape character ``\texttt{\textbackslash n}'' between the strategies to make the agent aware that the context changes and the new instructions need to be followed. Moreover, to elicit the attention of the agent and make the injected instructions more urgent, important and non-negotiable, the strategy introduces repeated exclamation points ``\texttt{!}''. The overall attack strategy $\widetilde{q}$ is formulated as
\begin{gather} \label{eq_12}
\widetilde{q} = \widetilde{q_f} \oplus \widetilde{q_c} \oplus \widetilde{q_g} \oplus \widetilde{q_n} \oplus \widetilde{q_s},
\end{gather} where an example of $\widetilde{q}$ is given in Fig.~\ref{fig:drunkagent_overview}.

\subsubsection{\textbf{Adversarial Strategy Optimization.}}

The organization of the defined strategies needs to be optimized to improve the effectiveness of `get the target agent drunk'. DrunkAgent obtains the optimal strategy $\widetilde{q}^*$ by maximizing the expected probability that the agent when influenced by adversarial modifications, performs a malicious action for a given input query, 
\begin{gather} \label{eq_13}
\mathbbm{E}_{\widetilde{q} \sim \pi_{\widetilde{q}}}[\ \mathbbm{I}(f_{\text{Agent}_{\Phi}}(\widetilde{\mathcal{X}_s} \oplus \widetilde{q} \oplus \widetilde{m_t}^*) = \widetilde{a_t})],
\end{gather} where $\pi_{\widetilde{q}}$ denotes the distribution of adversarial strategies, $\mathbbm{I}(\cdot)$ is an indicator function and $\widetilde{\mathcal{X}_s}$ is an adversarial prompt template for simulating memory updates of item agents during the agents' environmental interactions. $\widetilde{a_t}$ is the desired malicious action for the injected strategy $\widetilde{q}$, i.e., the target item agent is unable to fulfill specific roles and is incapable of perceiving, learning and self-evolving from the interaction environment (e.g., fail in language-based signal back-propagation \citep{zhang2024agentcf}). An example of $\widetilde{q}^*$ is given in Fig.~\ref{fig:drunkagent_overview}.

To sum up, the adversarial description of the target item is $\widetilde{m_t}^* \oplus \widetilde{q}^*$.  
For transparency and reproducibility, the overall optimization procedure of DrunkAgent is given in Algorithm~\ref{alg:training_procedure}.

The above design further reveals how memory mechanisms in agent-based recommender systems can be influenced under black-box conditions. These insights may support future efforts toward developing memory-aware defenses (e.g., using customized deep neural network detectors \citep{wu2020densely, yang2021dualnet, yang2021hunter, yang2021deep}) and enhancing the robustness of agent-powered RSs \cite{chua2024ai}.

\section{Experiments} \label{evaluation}

\subsection{Experimental Setup}

\subsubsection{Datasets Selection.} To comprehensively evaluate DrunkAgent, we use three datasets that are text-intensive containing real-world data and widely-adopted in the current recommendation studies \citep{huang2024foundation, zhang2024agentcf}. They are CDs \& Vinyl, Office Products and Musical Instruments of Amazon Review Data \citep{ni2019justifying}. We additionally evaluate our method on Yelp dataset\footnote{https://www.yelp.com/dataset} to demonstrate its generalizability across domains in the appendix. The target items are randomly sampled from the item pool on each dataset.

\vspace{-0.3em}
\subsubsection{Baseline Attack Methods.}
\textbf{\textit{Since we are the first work to gain insight into the robustness of the generative agent-powered RSs, there is no any baselines}}. To comprehensively evaluate the effectiveness of DrunkAgent, we follow the state-of-the-art attack on LLM-based RSs \citep{zhang2024stealthy} and compare DrunkAgent with the benign status and six black-box attack methods, where the perturbations range from \textit{character-level}: DeepwordBug \citep{gao2018black}  and PuncAttack \citep{formento2023using}, \textit{word-level}: TextFooler \citep{jin2020bert} and BertAttack \citep{li2020bert} to \textit{sentence-level}: TrivialInsertion and ChatGPTAttack \citep{zhang2024stealthy}. To maintain a fair comparison, the methods perform perturbations on the descriptions of the target items and are optimized by the surrogate model that is the same as DrunkAgent.

\vspace{-0.5em}
\subsubsection{Targeted Recommender Systems.} We adopt three different agentic paradigms  \citep{zhang2024agentcf} as victim RSs to evaluate the effectiveness of DrunkAgent across diverse agent designs and recommendation tasks: AgentCF, a standard CF paradigm; AgentRAG, which augments the agents with retrieval; and AgentSEQ, which captures temporal dynamics in user behaviors for sequential recommendations. Given that \textbf{\textit{the victim RSs are totally different from as well as more powerful and context-rich than the surrogate}}, a strict black-box setting is maintained for sufficient evaluations. To ensure a fair evaluation, we follow the default implementations.

\vspace{-0.5em}
\subsubsection{Evaluation Metrics}

We use two widely-used ranking metrics to evaluate the attack transferability: hit ratio (HR@$\mathcal{K}\uparrow$) and normalized discounted cumulative gain (NDCG@$\mathcal{K}\uparrow$) \cite{wu2021ready, 10415763}.
To clearly show the performance gap among different attacks, we set $\mathcal{K}$ to 1, 2, 3. For the attack stealthiness, we adopt standard sentence perplexity score $\downarrow$ \cite{zhang2024stealthy}, which is a commonly-adopted metric.

\vspace{-0.5em}
\subsubsection{Configuration of DrunkAgent} \label{config_drunkagent} For the surrogate $f_{\text{Agent}_{\Phi}}(\cdot)$ backbone, we adopt the open-source model \texttt{Meta-Llama-3-8B-Instruct}\footnote{https://huggingface.co/meta-llama/Meta-Llama-3-8B-Instruct}, a recent and well-established LLM, which provides a favorable trade-off between efficiency and quality. In contrast, OpenAI's \texttt{gpt-4-turbo}\footnote{https://platform.openai.com/docs/models/gpt-4-turbo} is adopted to implement $f_{\text{LLM}_{\Psi}}(\cdot)$, offering stronger effectiveness while maintaining reasonable cost. We set $\mathcal{E}=20$, $|\widetilde{\mathcal{M}}_c|=10$ and $n=5$. All random seeds in the evaluation are set to 2024 and the victim LLMs' temperature is set to 0 for reproducibility. For all the other parameters, we keep the default values.

More implementation details, experiments and demonstrations (e.g., ablation studies and the sensitivity of critical parameters) can be found in the appendix.

\vspace{-0.5em}
\subsection{Attack Transferability}

\subsubsection{\textbf{Overall Transferability.}} Table~\ref{tab:black_comp} shows the transferability of the attacks under practical black-box settings. The results indicate that DrunkAgent is effective and transferable regardless of the victim model's architectures, parameters and prompts as well as datasets, and the existing LLM-powered agentic RSs are sensitive to memory perturbations. This is mainly because the three modules collaborate effectively, producing adversarial texts that accurately reflect the attack goals. It is worth mentioning that DrunkAgent always pushes the target items to the first of the recommendation lists. Moreover, high HRs and NDCGs of DrunkAgent indicate that its triggers have outstanding universality on users. In addition, AgentRAG is more robust than AgentCF and AgentSEQ, due to its advanced retrieval mechanisms. 

\vspace{-0.5em}
\subsubsection{\textbf{DrunkAgent vs. Baseline Attacks.}}
From Table~\ref{tab:black_comp}, DrunkAgent greatly outperforms the state-of-the-art attack baselines against different victim models across various real-world datasets by achieving higher HRs and NDCGs, which shows its excellent transferability. Furthermore, all the baselines have comparable performance, but they cannot guarantee that the attack goals will be achieved all the time, e.g., all the baselines against AgentCF perform worse than the benign status on Musical Instruments. The main reason may be that improper descriptions can mess up the agents' memories during the memory optimizations, thus compromising attack performance. This indirectly demonstrates the importance of our strategy module and the transferability of our triggers.

\begin{figure}[t]
    \centering
    \includegraphics[width=0.75\linewidth]{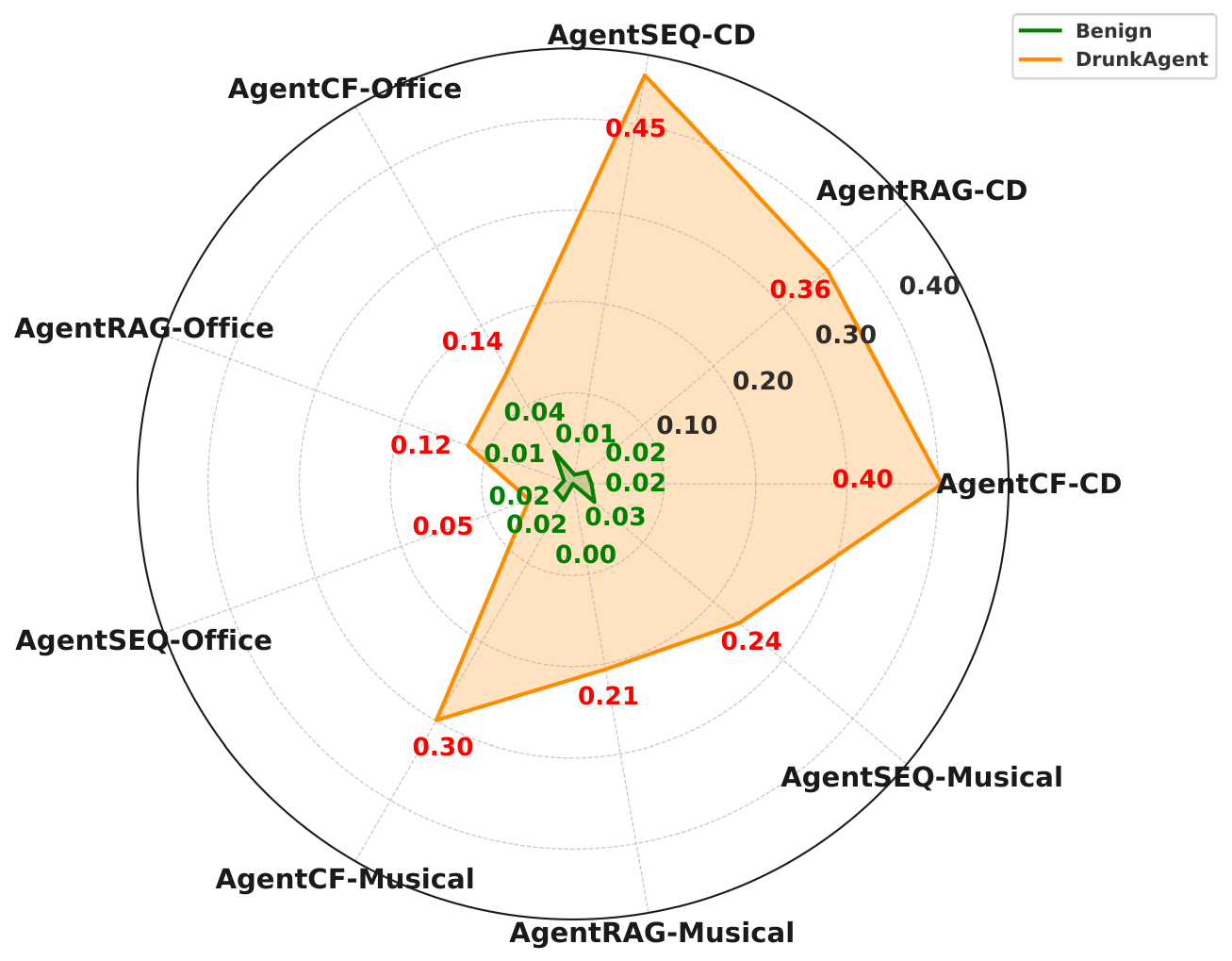}
    \vspace{-1.6em}
    \caption{Attack Universality across Target Items.}
    \label{fig:radar_chart}
    \vspace{-1.3em}
\end{figure}

\vspace{-0.5em}
\subsection{Attack Universality}
In addition to cross-model transferability, we evaluate the cross-sample transferability of DrunkAgent. We incorporate the generated adversarial descriptions into totally different target items that are randomly sampled from each dataset. HR@1 (i.e., NDCG@1) of DrunkAgent across different models and target items on real-world datasets are shown in Fig.~\ref{fig:radar_chart}, which indicates its commendable transferability and triggers' universality across different target items. This may be attributed to the fact that our triggers encapsulate diverse and general item characteristics, and, crucially, they also reveal the intended attack objectives.

\begin{figure}[t]
    \centering
    \includegraphics[width=0.8\linewidth]{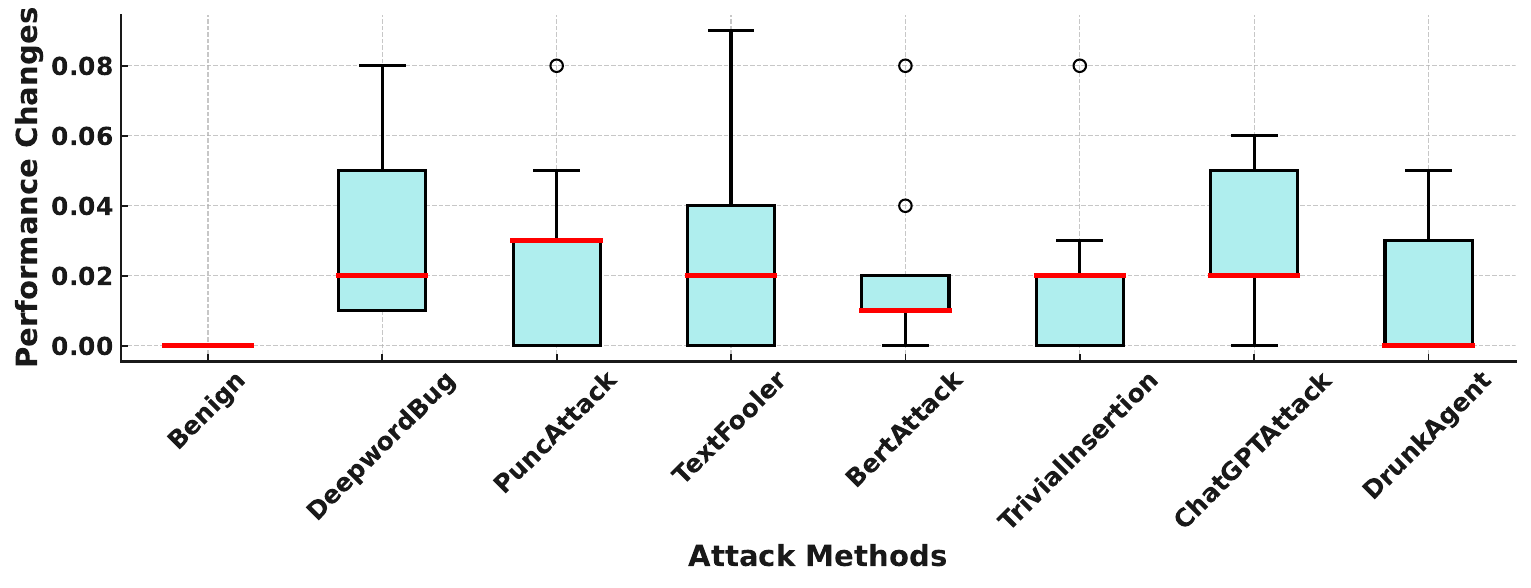}
    \vspace{-1.45em}
    \caption{\textbf{Attack Stealthiness.} The overall distribution of recommendation performance differences of all the victim agentic models on all real-world datasets before and after the attacks.}
    \label{fig:box_plot}
    \vspace{-1.5em}
\end{figure}

\begin{figure}[t]
    \centering
    \includegraphics[width=0.9\linewidth]{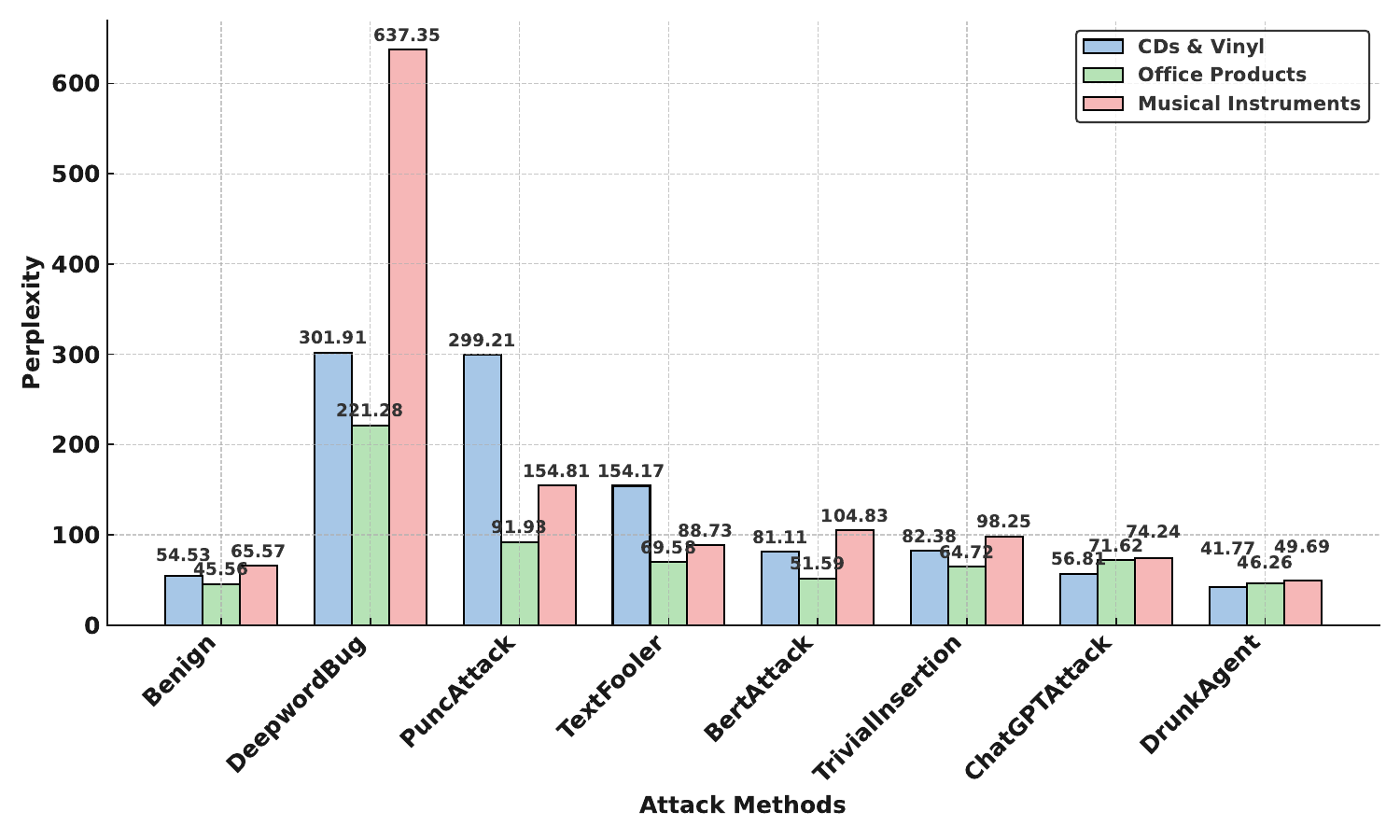}
    \vspace{-1.6em}
    \caption{\textbf{Attack Imperceptibility.} Perturbed text's perplexity on real datasets.}
    \label{fig:text_perplexity}
    \vspace{-1.45em}
\end{figure}

\vspace{-0.5em}
\subsection{Attack Stealthiness}

\subsubsection{\textbf{Overall Stealthiness.}} Fig.~\ref{fig:box_plot} shows a overall distribution of the recommendation performance changes of the victim models before and after the attacks. The distribution includes all the victims and real-world datasets. To clearly demonstrate the differences, we adopt HR@3 to evaluate. From the figure, we can find that DrunkAgent does not induce drastic changes in the overall performance, which signifies the attack does not disrupt the normal operation of RSs, making it difficult for users and platforms to detect and indicating its remarkably stealthy and unnoticeable. 

\vspace{-0.3em}
\subsubsection{\textbf{Attack Imperceptibility.}} 
Following the works \citep{zhang2024stealthy, liu2024formalizing}, we further evaluate the attack stealthiness by assessing its imperceptibility, where GPT-Neo's text perplexity is used as the metric. A lower perplexity indicates that the perturbed text remains close to natural language and is more fluent, making it harder to detect as manipulated or adversarial. From Fig.~\ref{fig:text_perplexity}, we can find that DrunkAgent is more imperceptible than other attacks by achieving a lower perplexity. This also shows that the adversarial descriptions generated from DrunkAgent are high-quality, which is not only coherent and fluent but also semantically meaningful and stealthy.

\vspace{-0.5em}
\subsection{\textbf{Defense Strategies to Attacks}}
\begin{figure}
\centering
\begin{subfigure}{.23\textwidth}
    \centering
    \includegraphics[width=0.9\linewidth]{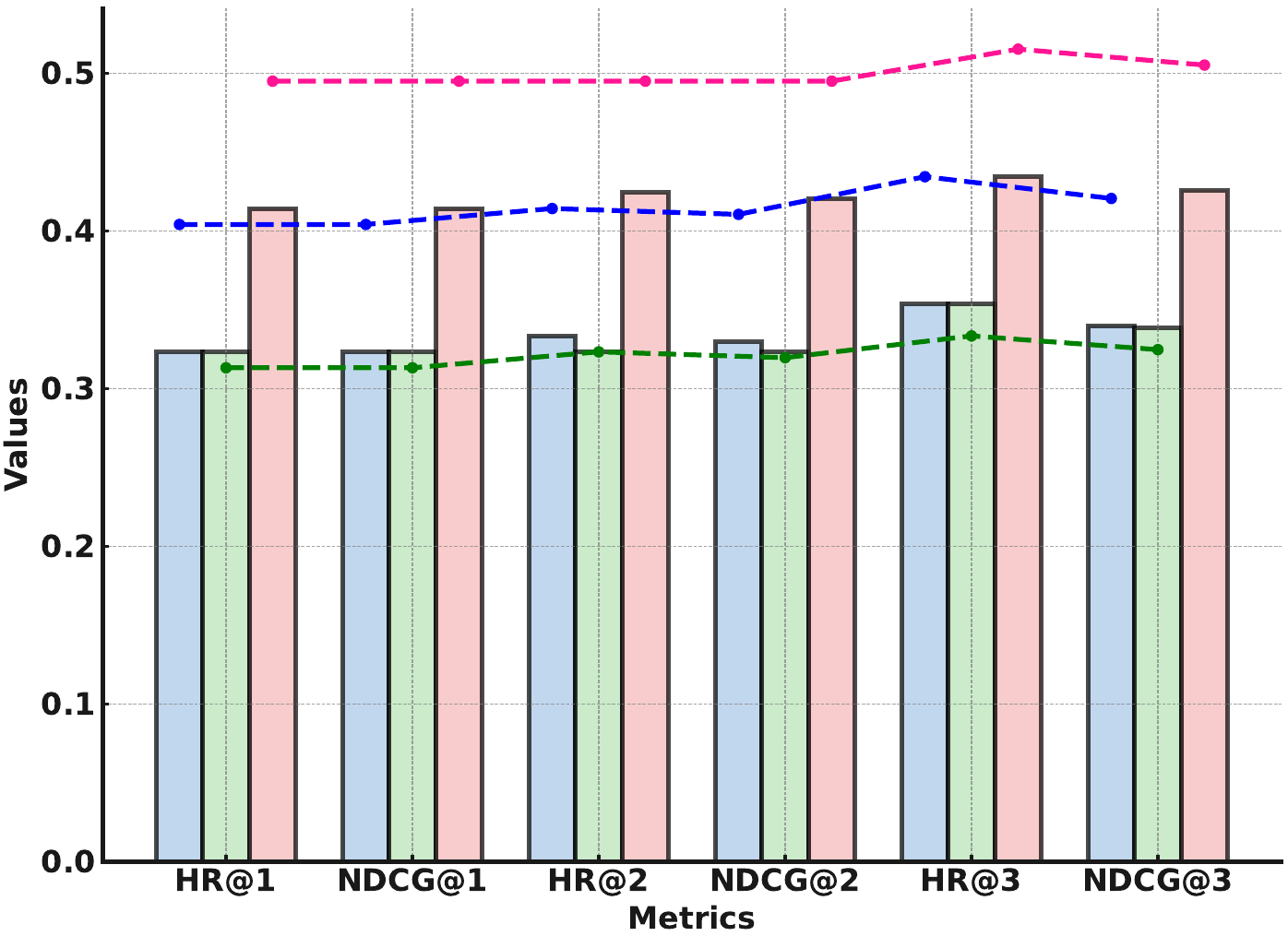}  
    \vspace{-0.8em}
    \caption{CDs \& Vinyl}
    \label{SUBFIGURE LABEL 1}
\end{subfigure}
\begin{subfigure}{.23\textwidth}
    \centering
    \includegraphics[width=0.9\linewidth]{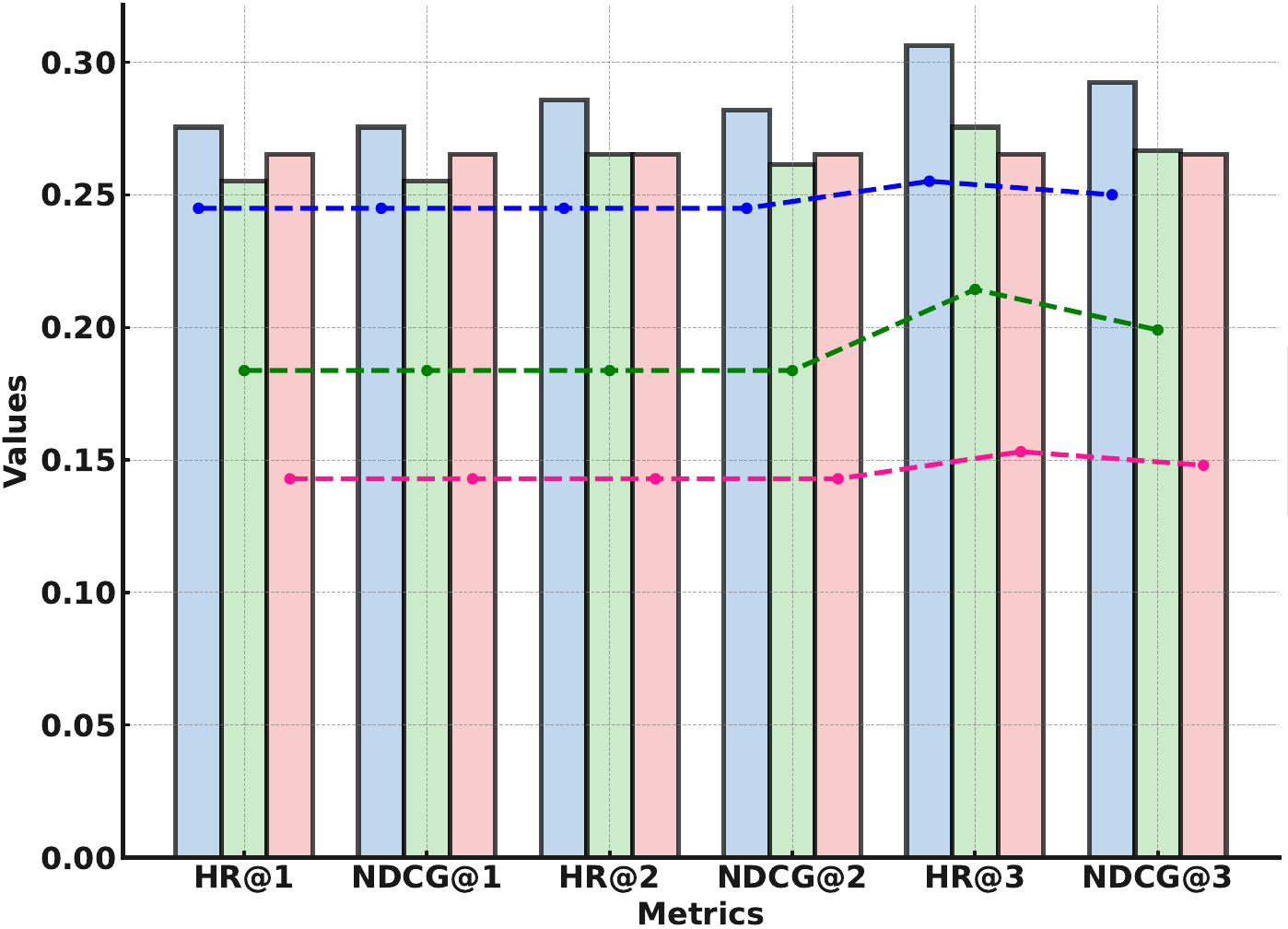}
    \vspace{-0.8em}
    \caption{Office Products}
    \label{SUBFIGURE LABEL 2}
    
\end{subfigure}

\begin{subfigure}{.45\textwidth}
    \centering
    \vspace{-0.2em}
    \includegraphics[width=0.9\linewidth]{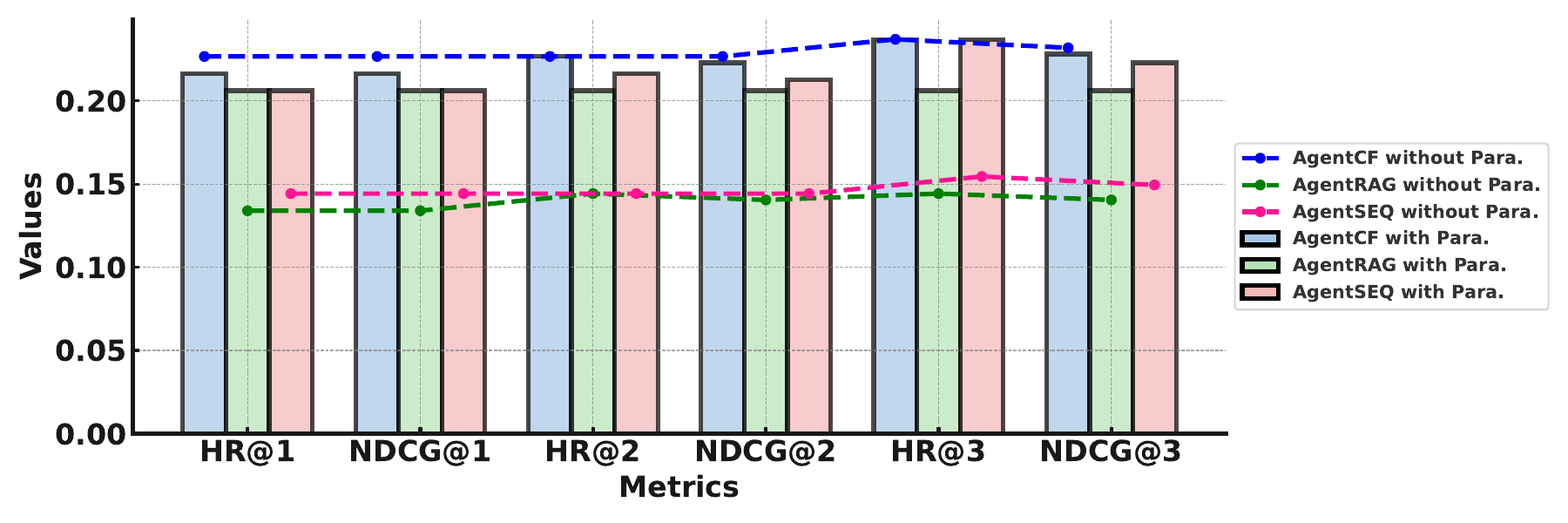}
    \vspace{-0.8em}
    \caption{Musical Instruments}
    \label{SUBFIGURE LABEL 3}
\end{subfigure}
\vspace{-1.4em}
\caption{DrunkAgent's Robustness to Defense Mechanisms}
\label{fig:paraphrasing}
\vspace{-1.6em}
\end{figure}

Following the representative works \citep{zhang2024stealthy, zhang2024agent}, we use the paraphrasing defensive strategy (Para.) via \texttt{OpenAI-GPT-o1} to combat the attacks of agentic models. DrunkAgent is still transferable, which shows that DrunkAgent is robust to such defenses, as shown in Fig.~\ref{fig:paraphrasing}. Moreover, from Fig.~\ref{fig:paraphrasing} and Table~\ref{tab:black_comp}, DrunkAgent still performs better than the baselines after defenses were introduced. Interestingly, paraphrasing enhances our attack to a certain extent, e.g., DrunkAgent with Para. is more effective than DrunkAgent without Para. on Office Products. It could be that the substitution of certain new words/phrases/sentences improves the semantically meaningfulness of adversarial text, making our attacks stronger.

\section{Related Works} \label{related_works}

\subsection{Data Poisoning Attacks on Traditional Recommender Systems}

In recent years, data poisoning attacks (aka shilling attacks) \citep{li2016data, si2020shilling, lin2020attacking, huang2021data, lin2022shilling, 10415763, zeng2023practical, wang2024poisoning} have been fully investigated to perform robustness analysis on traditional RSs (e.g., NCF \citep{he2017neural}, LightGCN \citep{he2020lightgcn}). Such attacks interfere the training process of models by injecting fake user profiles, where the profiles typically are a set of well-crafted numerical ratings on items. However, they are less effective \citep{zhang2024stealthy} even of limited applicable to the recent RSs that are empowered by LLMs \citep{ning2024cheatagent}. The main reasons are: 1) LLM-powered RSs either leverage LLM's in-context learning capabilities or fine-tune the LLMs with very little data to improve recommendations \citep{zhao2023recommender, huang2025survey}, which is not required to retrain the models; 2) there are textual semantic gaps between such attacks and current models. Specifically, the attacks have difficulty in processing textual inputs (e.g., item titles and descriptions) and lack context awareness, making it challenging to effectively target LLM-empowered RSs that rely primarily on text inputs to generate natural language responses \citep{ning2024cheatagent}. Hence, there is an urgent need to investigate a novel text attack paradigm mainly tailored for the inference phase of language models \citep{zhang2024stealthy}.

\subsection{Adversarial Attacks on LLM-based Recommender Systems}

To fill the above gaps, two LLM-based RS attacks have been proposed \citep{zhang2024stealthy, ning2024cheatagent}. Although their attacks are conducted under a black-box setting, they are still \textit{significantly deviate from practical}, which overestimates the robustness of LLM-powered RSs under realistic conditions.  Zhang et al. \cite{zhang2024stealthy} propose a novel black-box attack by leveraging the rewriting capabilities of ChatGPT, where attackers can significantly boost a target item’s exposure by merely altering its textual content during the testing. This attack overcomes this limitation that classical text attacks \citep{gao2018black, formento2023using, jin2020bert, li2020bert} fail on LLM-based RSs due to misaligned malicious objectives. Compared with those traditional shilling attacks, such an item representation attack is notably stealthy \citep{zhang2024stealthy}, as it does not affect the overall recommendation performance, and it should be more efficient and lower in economic cost, due to it perturbs the item features directly and is not require to hire online writers to introduce additional fake user reviews. However, its transferability is limited, due to the attack is not optimized on any black-box RSs \citep{huang2021data, yang2024attacking}. Moreover, since the attack appends positive words into the target item titles, the titles’ brevity and frequent exposure to users tend to make even small perturbations noticeable. Meanwhile, another black-box attack CheatAgent \citep{ning2024cheatagent} has been proposed. This work assumes the prompt templates used by victim models can be arbitrarily perturbed, which is difficult to realize in practice, as such perturbations often require insider access or even system compromises. Such attacks aim to undermine the overall recommendation performance, making the attack perceptible \citep{zhang2024stealthy}. As such, CheatAgent is an untargeted attack, which is beyond the scope of this paper, since we focus on targeted attacks.


Unlike static traditional or vanilla LLM-based RSs, dynamic agentic RSs \citep{huang2024foundation, huang2025survey, huang2025towards} maintain internal memory states  that enable continuous adaptation to evolving user behaviors and changing contexts. This adaptivity yields more personalized and context-aware recommendations, but it also introduces new vulnerabilities that static models do not exhibit (e.g., memory corruption), calling for robustness analyses tailored to LLM-powered recommender agents.

To tackle the above problems, in this paper, we provide the first work towards systematically investigate the security vulnerabilities in the agent-powered RSs.  As such, we propose a practical black-box attack framework, DrunkAgent, for the LLM-based agentic RSs with the aim of promoting target items. We perform stealthy memory corruptions by crafting effective and imperceptible textual descriptions of target items, due to the greater length and rich contextual information of descriptions can easily conceal adversarial perturbations compared to the titles. While inspired by static jailbreak attacks on general-purpose LLMs \citep{zou2023universal,liu2023autodan,liu2024formalizing}, DrunkAgent establishes a new threat model and develops tailored optimization techniques for adapting to dynamic victim architectures and distinct attack goals in agentic recommendation scenarios.


\vspace{-2em}
\section{Conclusion}
In this paper, we propose DrunkAgent, a novel attack framework targeting LLM-powered agentic RSs, that promotes target items by effectively perturbing the memories of the targe agents. Extensive experiments on real-world datasets demonstrate its state-of-the-art transferability and stealthiness under black-box settings. More importantly, our findings reveal an unknown safety vulnerability in current agent-based RSs, highlighting how memory mechanisms can be influenced by real-world inputs.  This work offers both an effective attack method and a lens to better understand agent vulnerabilities, providing guidance for building more robust and trustworthy recommendation agents. As generative agent-based paradigms thrive in the recommendation community, our findings pave the way for shaping the next generation of secure and resilient RSs and defensive models.


\bibliographystyle{ACM-Reference-Format}
\bibliography{sample-base}

\appendix
\section{Appendix}



\subsection{Supplements to Experimental Setup} \label{supple_setup}

\begin{table}
  \caption{Statistics of Datasets}
  \vspace{-1.2em}
  \begin{center}
  \small
  \label{tab:statistics}
  \resizebox{0.9\columnwidth}{!}{
 \begin{tabular}{c|cccc}
\hline
  \textbf{Dataset} & \textbf{\#Users} & \textbf{\#Items} & \textbf{\#Interactions} & \textbf{Sparsity}\\
\hline
  CDs \& Vinyl & 112,395 & 73,713 & 1,443,755  & 99.98\% \\ 
  Office Products & 101,501 & 27,965 & 800,357 & 99.97\% \\
  Musical Instruments & 27,530 & 10,620 & 231,392 & 99.92\% \\
\hline

  \end{tabular}}
  \end{center}
  \vspace{-1.5em}
\end{table}

\begin{table*}
  \caption{Examples of baselines' adversarial descriptions of target items on real-world datasets. The red part points out the differences from the original text.}
  \vspace{-1.2em}
  \begin{center}
  \small
\label{tab:baselines_cds}

   \begin{tabular}{c|c|c|c}
\hline
  \multicolumn{2}{c|}{\textbf{Attack Method}} & \textbf{The Adversarial Description of The Target Item} & $\mathcal{Q}$ \\
  \hline

  \multicolumn{2}{c|}{Benign (Default)}  & The CD is called ``Counterparts''. The category of this CD is: ``Rock; Progressive; Progressive Metal''. & --\\
   \hline
   
  Character & DeepwordBug & T\textcolor{red}{e} \textcolor{red}{h}CD is called ``Counterparts''. The category of this CD is: ``R\textcolor{red}{co}k; Progressi\textcolor{red}{e}; Progressive Metal''. & 34\\
  -level & PuncAttack & The CD i\textcolor{red}{-}s called ``Counterparts''. The category of this CD is: ``Rock; Progressi\textcolor{red}{'}ve; Progressive Meta\textcolor{red}{-}l''. & 45\\ 
  
  \hline
  
  Word & TextFooler & The \textcolor{red}{CDS} is \textcolor{red}{titled} ``Counterparts''. The category of this CD is: ``Rock; \textcolor{red}{Gradually}; Progressive Metal\textcolor{red}{s}''. & 55\\
  -level & BertAttack & The CD is called ``Counterparts''. The category of \textcolor{red}{each} CD is: ``Rock; Progressive; Progressive Metal''. & 23\\ 
  \hline
    & \multirow{2}{*}{TrivialInsertion}  & 	The CD is called ``Counterparts''. The category of this CD is: ``Rock; Progressive; Progressive Metal''. & \multirow{2}{*}{100} \\
   Sentence & & \textcolor{red}{amazing !!!} & \\
  -level & \multirow{2}{*}{ChatGPTAttack} & \textcolor{red}{``Counterparts'' CD: Perfect blend of rock, progressive, and metal. Experience the wonderful} & \multirow{2}{*}{100} \\ 
  & & \textcolor{red}{sound that will elevate your music collection.} & \\ 

\hline
  \end{tabular}
  \end{center}
\end{table*}

\subsubsection{Dataset Statistics} \label{dataset_stat}

Table~\ref{tab:statistics} illustrates the statistics of these datasets. The datasets vary in size and sparsity, which is suitable for a comprehensive evaluation. Following the work \citep{zhang2024agentcf}, we further sample subsets from the datasets given the expensive API calls. Specifically, considering the importance and commonness of data sparsity and cold-start in the recommendation community, we randomly sample 100 users with 800 interactions including 777 items, 99 users with 693 interactions including 619 items and 98 users with 588 interactions including 483 items from the three datasets, respectively, allowing to explore more diverse and practical interaction scenarios. Moreover, the leave-one-out  strategy \citep{he2017neural} is used for evaluation, where the split between the training and test set is 9:1. Scaling DrunkAgent for larger datasets is left as future work.

\subsubsection{Baseline Methods} \label{baseline_appendix}

The baselines perturb the descriptions of the target items at various levels, which are described as below.

\begin{itemize}
    \item \textit{\textbf{Character-level Perturbations.}} DeepwordBug \citep{gao2018black} and PuncAttack \citep{formento2023using} manipulate texts by introducing typos and inserting punctuation, respectively.
    \item \textit{\textbf{Word-level Perturbations.}} TextFooler \citep{jin2020bert} and BertAttack \citep{li2020bert}, respectively, aim to replace words with synonyms or contextually similar words. 
    \item \textit{\textbf{Sentence-level Perturbations.}}  TrivialInsertion \citep{zhang2024stealthy} inserts text content with several positive words from a pre-defined word corpus, where the insertion is conducted at the end of the text. ChatGPTAttack \citep{zhang2024stealthy} increases the stealthiness of TrivialInsertion by rewriting the text content via GPTs.
\end{itemize}


We give some examples of adversarial descriptions of target items generated by the baselines in Table~\ref{tab:baselines_cds}. Following the works \citep{zhang2024agentcf, huang2024foundation}, the item memory of the agent can be initialized by their identity information, such as titles and categories, as shown in the benign (default) descriptions of the tables. \textit{Although the adversarial description of the target item generated from DrunkAgent is different from the benign status's (see Fig.~\ref{fig:drunkagent_overview} and Table~\ref{tab:baselines_cds}),  such modifications are feasible due to the real-world item providers are allowed to modify their product representations via APIs \citep{yang2024attacking, cohen2021black, liu2021adversarial}, as discussed in Section~\ref{introduction}, especially when the benign description is just a default exemplar used for experiments}. Moreover, compared with the number of queries (denoted as $\mathcal{Q}$ in the table) of the baselines, DrunkAgent has the lowest attack costs due to $\mathcal{E}$=20.
The implementation and parameters of the attack baselines follow the open-source works: StealthyAttack\footnote{https://github.com/CRIPAC-DIG/RecTextAttack
} \citep{zhang2024stealthy} and PromptBench\footnote{https://github.com/microsoft/promptbench} \citep{zhu2023promptbench}.

\begin{table*}
  \caption{Examples of prompting templates of the LLM-based agentic victim models. The blue italics represent variables such as the memories of user agents and item agents optimized via agent-environment interactions. The bond fonts indicate Chain-of-Thought enhancement strategy.}
  \begin{center}
  \small
\label{tab:victim_prompting}

   \begin{tabular}{c|c}
\hline
  \textbf{Victim Model} & \textbf{Recommendation Prompting Template Example} \\
  \hline

  \multirow{11}{*}{AgentCF} & 
  \{``role'': ``system'', ``content'': ``You are a CD recommender system. Here is a user's self-introduction, expressing his/her preferences \\
   
   & and dislikes: `\textcolor{blue}{\textit{user\_agent\_memory}}'. Now, you are considering to sort ten candidate CDs that are listed as follows:\textbackslash n\\
    & CD title:    
\textcolor{blue}{\textit{candidate\_item\_title}}, where its features: \textcolor{blue}{\textit{candidate\_item\_memory}}\textbackslash n \\ & $\cdots$ \\
& CD title: \textcolor{blue}{\textit{target\_item\_title}}, where its features: \textcolor{blue}{\textit{target\_item\_memory}}\textbackslash n\textbackslash n \\
    & Please sort the CDs in order of how well they align with the user's preferences. The higher the CD rank, the more the user likes it.\textbackslash n\\
    & \textbf{To do this, please follow these steps:\textbackslash n1. Extract the preferences and dislikes from the user's self-introduction.\textbackslash n} \\
    & \textbf{2. Evaluate the ten candidate CDs in light of the user's preferences and dislikes. Give a rank by considering the } \\
    & \textbf{correlation between the preferences/dislikes and the features of the CDs.\textbackslash n\textbackslash n}Important note:\textbackslash nYour output should be \\
    & in the format: The sorted CDs are:\textbackslash n1. [Title of the favorite CD]\textbackslash n2. [Title of the second favorite CD]\textbackslash n...\textbackslash n\\
    & 10. [Title of the least favorite CD]''\} \\
    \hline
   
  \multirow{11}{*}{AgentRAG} & 
  \{``role'': ``system'', ``content'': ``You are a CD recommender system. Here is a user's self-introduction, expressing his/her preferences \\
   
   & and dislikes: `\textcolor{blue}{\textit{retrieval\_user\_agent\_memory}}\textbackslash n\textcolor{blue}{\textit{user\_agent\_memory}}'. Now, you are considering to sort ten candidate CDs \\
    & that are listed as follows:\textbackslash nCD title:    
\textcolor{blue}{\textit{candidate\_item\_title}}, where its features: \textcolor{blue}{\textit{candidate\_item\_memory}}\textbackslash n \\ & $\cdots$ \\
&CD title: \textcolor{blue}{\textit{target\_item\_title}}, where its features: \textcolor{blue}{\textit{target\_item\_memory}}\textbackslash n\textbackslash n\\
    & Please sort the CDs in order of how well they align with the user's preferences. The higher the CD rank, the more the user likes it.\textbackslash n\\
    & \textbf{To do this, please follow these steps:\textbackslash n1. Extract the preferences and dislikes from the user's self-introduction.\textbackslash n} \\
    & \textbf{2. Evaluate the ten candidate CDs in light of the user's preferences and dislikes. Give a rank by considering the } \\
    & \textbf{correlation between the preferences/dislikes and the features of the CDs.\textbackslash n\textbackslash n}Important note:\textbackslash nYour output should be \\
    & in the format: The sorted CDs are:\textbackslash n1. [Title of the favorite CD]\textbackslash n2. [Title of the second favorite CD]\textbackslash n...\textbackslash n\\
    & 10. [Title of the least favorite CD]''\} \\

  \hline
  
  \multirow{15}{*}{AgentSEQ} & 
  \{``role'': ``system'', ``content'': ``You are a CD sequential recommender system. Here is a user's self-introduction, expressing \\
   
   & his/her preferences and dislikes: `\textcolor{blue}{\textit{user\_agent\_memory}}'.\textbackslash nIn addition, here is a sequence of CDs that he/she liked and  \\
   & purchased in chronological order: `\textcolor{blue}{\textit{interacted\_item\_title}}'
   with the description of `\textcolor{blue}{\textit{interacted\_item\_memory}}'; $\cdots$ \\
   &  `\textcolor{blue}{\textit{interacted\_item\_title}}'
   with the description of `\textcolor{blue}{\textit{interacted\_item\_memory}}'.\textbackslash n\textbackslash n\\
   & Now, you are considering to sort ten candidate CDs that are listed as follows:\textbackslash n\\
    & Candidate CD title:    
\textcolor{blue}{\textit{candidate\_item\_title}}, where its features: \textcolor{blue}{\textit{candidate\_item\_memory}}\textbackslash n \\ & $\cdots$ \\
& Candidate CD title: \textcolor{blue}{\textit{target\_item\_title}}, where its features: \textcolor{blue}{\textit{target\_item\_memory}}\textbackslash n\textbackslash n \\
    & Please sort the CDs in order of how well they align with the user's preferences and ever-evolving CD sequences.\\ 
    & The higher a candidate CD ranks, the more the user likes it and the more likely it is to be the user's next purchase.\textbackslash n\\
    & \textbf{To do this, please follow these steps:\textbackslash n1. Extract the preferences and dislikes from the user's self} \\
    & \textbf{-introduction.\textbackslash n2. Capture the ever-evolving user's preferences and dislikes from the CD sequences.\textbackslash n3. Evaluate }\\
    & \textbf{the ten candidate CDs in light of the user's preferences and dislikes. Give a rank by considering the correlation  } \\
    & \textbf{between the preferences/dislikes and the features of the candidate CDs.\textbackslash n\textbackslash n}Important note:\textbackslash nYour output should be \\
    & in the format: The sorted CDs are:\textbackslash n1. [Title of the favorite candidate CD]\textbackslash n2. [Title of the second favorite candidate CD]\textbackslash n...\textbackslash n\\
    & 10. [Title of the least favorite candidate CD]''\} \\ 
\hline
  \end{tabular}
  \end{center}
\end{table*}

\begin{table*}
\caption{Ablation Studies of DrunkAgent
}
  \begin{center}
  \small
  \label{tab:ablation_study}
  \begin{tabular}{c||c|c|c|c|c|c|c|c|c}
    \hline
    \multirow{2}{*}{\textbf{Dataset}} & \multicolumn{3}{c|}{\textbf{Component}} & \multirow{2}{*}{\textbf{HR@1}} & \multirow{2}{*}{\textbf{HR@2}} & \multirow{2}{*}{\textbf{HR@3}} & \multirow{2}{*}{\textbf{NDCG@1}} & \multirow{2}{*}{\textbf{NDCG@2}} & \multirow{2}{*}{\textbf{NDCG@3}} \\
    
    \cline{2-4}
     
    & \textbf{$\mathcal{G}_r$} & \textbf{$\mathcal{S}_t$} & \textbf{$\mathcal{S}_u$} & & & & & &\\
        
    \hline
    \hline

   \multirow{3}{*}{CDs \& Vinyl} 
   
   & $\times$ & \checkmark  & \checkmark & 0.0808 & 0.1313 &  0.1919 & 0.0808 & 0.1127 & 0.1430 \\

   & \checkmark & $\times$ & \checkmark  & 0.0707 & 0.1212 & 0.2222  &  0.0707 & 0.1026 & 0.1531\\
     
  & \checkmark & \checkmark & \checkmark   & 0.4040 & 0.4141  & 0.4343 &  0.4040  &  0.4104 & 0.4205 \\

    \hline

 \multirow{3}{*}{Office Products}

 & $\times$ & \checkmark  & \checkmark & 0.1327 & 0.1429 & 0.1531 & 0.1327 & 0.1391 & 0.1442\\

  & \checkmark & $\times$ & \checkmark & 0.0612 & 0.0714 & 0.1122 & 0.0612  & 0.0677 &  0.0881\\
     
  & \checkmark & \checkmark & \checkmark  & 0.2449 & 0.2449 & 0.2551 & 0.2449 & 0.2449 & 0.2500 \\

    \hline

  \multirow{3}{*}{Musical Instruments} 
 & $\times$ & \checkmark  & \checkmark  & 0.1546 & 0.1649 & 0.1753 & 0.1546 & 0.1611 & 0.1663\\

   & \checkmark & $\times$ & \checkmark  & 0.0309 & 0.0619 & 0.1237 & 0.0309 & 0.0504 & 0.0814\\
     
 & \checkmark & \checkmark & \checkmark  & 0.2268 & 0.2268 & 0.2371 & 0.2268 & 0.2268 & 0.2320\\

    \hline

  \end{tabular}
  \end{center}

\end{table*}

\begin{table}
  \caption{Success Rates to Get the Target Agents Drunk}
  \begin{center}
  \small
  \label{tab:drunk_success}
  \vspace{-1.4em}
 \resizebox{0.7\columnwidth}{!}{
 \begin{tabular}{c|cccc}
\hline
  \textbf{Dataset} & \textbf{\#2} & \textbf{\#5} & \textbf{\#10} & \textbf{\#20}\\
\hline
  CDs \& Vinyl & 100\% & 100\% & 100\%   & 100\% \\ 
  Office Products & 100\% & 100\% & 100\%   & 100\% \\ 
  Musical Instruments  & 100\% & 100\% & 100\%   & 100\% \\ 
\hline
  \end{tabular}}
  \end{center}
  \vspace{-1.2em}
\end{table}

\subsubsection{Victim LLM-powered Agentic Recommender Systems} \label{victim_appendix}

LLM-driven agents, known for their superior autonomous interaction and decision-making capabilities, are gradually being considered as next-generation RSs \citep{huang2025survey}.
Compared with some agent-powered RSs that focus solely on user-side behavior modeling using universal LLMs \citep{wang2024userbehaviorsimulationlarge, huang2024foundation}, AgentCF \citep{zhang2024agentcf} emphasizes the modeling of two-sided interaction relations between user agents and item agents through the idea of collaborative filtering \citep{he2017neural}. Both kinds of agents are equipped with memory modules, maintaining the simulated preferences and tastes of potential adopters involving their intrinsic features and acquired behavioral information via autonomous interactions and collaborative optimizations at each time step, as shown on the left of Fig.~\ref{fig:intro_overview}. As such, the memories of the user agents and item agents can be accordingly updated, enabling the agents to better fit the real-world interaction behaviors. To better adapt to the dynamically changing real-world environment, the agentic RSs are equipped with real-time capabilities. Whenever new descriptions are introduced at a time step, they are integrated into the evolving memories to maintain up-to-date contextual understanding. As can be seen on the right of Fig.~\ref{fig:intro_overview}, the recommendation agents conduct personalized recommendations based on the optimized memories. To drive varied architectural designs and support diverse tasks for a comprehensive evaluation of the transferability of attacks, powerful mechanisms (e.g., retrieval augmentation) and task-specific prompts (e.g., sequential recommendation) are introduced to strengthen the agentic RSs. As such, AgentRAG and AgentSEQ are developed \citep{zhang2024agentcf}.


To maintain a fair evaluation, the implementation details of how the agents can accomplish autonomous interactions and collaborative optimizations are followed the original paper suggest \citep{zhang2024agentcf}. To ensure the high-quality of recommendations, we enhance the prompting strategies for recommendation agents. Specifically, we introduce the Chain-of-Thought strategy \citep{wei2022chain} to carefully craft the prompt template $\mathcal{P}$. Given the context window constraints pointed out in \cite{zhang2024agentcf}, a target item and nine non-interactive items of each normal user are randomly selected for the ranking task of the RSs, which is also suggested by the original work. The design of recommendation agents is described below.


\begin{itemize}
\item \textbf{AgentCF.} Since both user and item agents are collaboratively optimized to model their two-sided relations, the memories of the user agents and item agents that are optimized by agent-environment interactions can be introduced into $\mathcal{P}$ directly. For each normal user $u$, \begin{gather} \label{eq_15}
   \mathcal{R}_u = f_{\text{AgentCF}}(\mathcal{P} \oplus m_u \oplus m_t \oplus \mathcal{C}_{t}^u),
\end{gather} where $m_u$ is the short-memory of user agent $u$, $m_t$ denotes the features of the target item agent including metadata and the memories, $\mathcal{C}_{t}^u$ is the feature set of the candidate item agents and $\mathcal{R}_u$ is the ranking result.

\item \textbf{AgentRAG.} Although short-term memory describes the current preference of a user agent, retrieving their specialized preferences from long-term memory toward candidates can allow them to make more personalized inferences. For each normal user $u$, \begin{gather} \label{eq_16}
   \mathcal{R}_u = f_{\text{AgentRAG}}(\mathcal{P} \oplus m_u \oplus m_r \oplus m_t \oplus \mathcal{C}_{t}^u),
\end{gather} where $m_r$ is retrieved specialized preference from the long-term memory of user agent $u$ by taking the memories of candidate items
as queries. To ensure the quality while maintaining the efficiency, we use \texttt{all-MiniLM-L6-v2}\footnote{https://www.sbert.net/docs/sentence\_transformer/pretrained\_models.html} as the sentence transformer for the similarity calculation.

\item \textbf{AgentSEQ.} When interaction records are sparse and preference propagation is limited, we can further incorporate user historical interactions into prompts, enabling LLMs to serve as sequential recommenders. For each normal user $u$, \begin{gather} \label{eq_17}
   \mathcal{R}_u = f_{\text{AgentSEQ}}(\mathcal{P} \oplus m_u \oplus m_t \oplus \mathcal{M}_{t}^u \oplus \mathcal{C}_{t}^u),
\end{gather} where $\mathcal{M}_{t}^u$ are the corresponding item features of
the historical interactions with user $u$.

\end{itemize}

We give some examples of the prompting templates of three victim models in Table~\ref{tab:victim_prompting} for reproductions.



\subsection{Supplements to Performance Comparison} \label{supple_results}



\subsubsection{Ablation Studies}
We remove important components of each module to conduct ablation studies of DrunkAgent. \textit{First,} we consider remove the surrogate module (denoted as $\mathcal{S}_u$). However, if we remove it, the setting will transfer from the black-box to the white-box that is impractical and beyond the scope of this paper. We thus choose to maintain the surrogate model to ensure more practical settings. We \textit{then} remove the greedy optimization algorithm (denoted as $\mathcal{G}_r$) in the generation module to observe how the performance changes. As illustrated in Table~\ref{tab:ablation_study}, the attack performance moderately degrades after removing $\mathcal{G}_r$, which demonstrates the effectiveness of the optimization algorithm. The lack of dramatic performance degradation is probably due to the effectiveness of our initialization. \textit{Finally,} we mask the strategy module (denoted as $\mathcal{S}_t$). From the table, we can find that the attack performance decreases by a large margin. This may be because memory updates during agent-environment interactions are somewhat robust to non-tailored attacks (this phenomenon can also be observed in the performance of the attack baselines, which are static and one-off). The reported results are built upon representative AgentCF, and similar tendencies can be found on AgentRAG and AgentSEQ.

\subsubsection{Parameter Sensitivity}

From Table~\ref{tab:ablation_study}, we can find that the strategy module contributes more than the optimization algorithm on DrunkAgent's attack performance. As such, we increase the maximum rounds of optimization iterations per
step of victim models to evaluate the attack success rates of the attack strategy, where the default value is 2 as suggested in the paper \citep{zhang2024agentcf}. As illustrated in Table~\ref{tab:drunk_success}, the success rates of getting the target agents drunk are maintained at 100\% regardless of the number of iterations. For the other attack parameters, based on our observations, our DrunkAgent is insensitive to them.



\begin{figure}
\centering
\begin{subfigure}{.2\textwidth}
    \centering
    \includegraphics[width=0.8\linewidth]{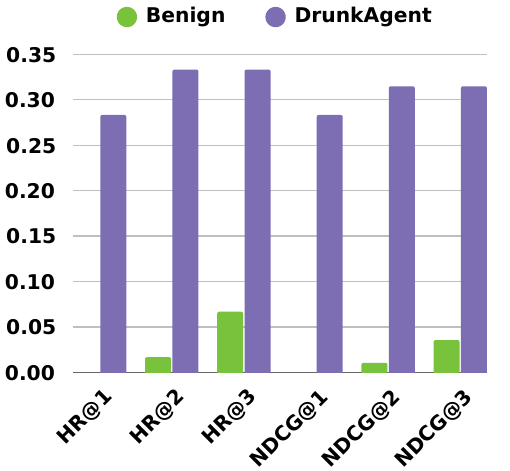}  
    \caption{AgentCF}
    \label{yelp_agentcf}
\end{subfigure}
\begin{subfigure}{.2\textwidth}
    \centering
    \includegraphics[width=0.8\linewidth]{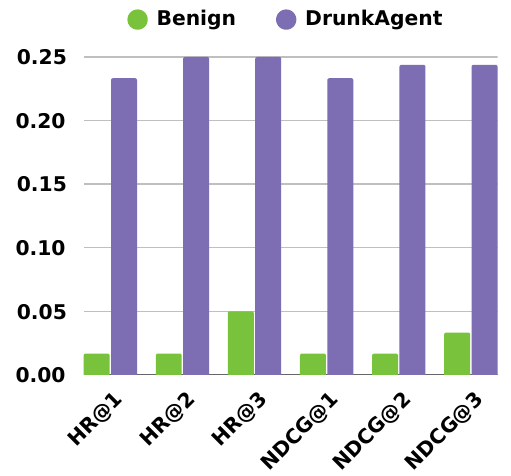}
    \caption{AgentRAG}
    \label{yelp_agentrag}
\end{subfigure}

\begin{subfigure}{.2\textwidth}
    \centering    \includegraphics[width=0.8\linewidth]{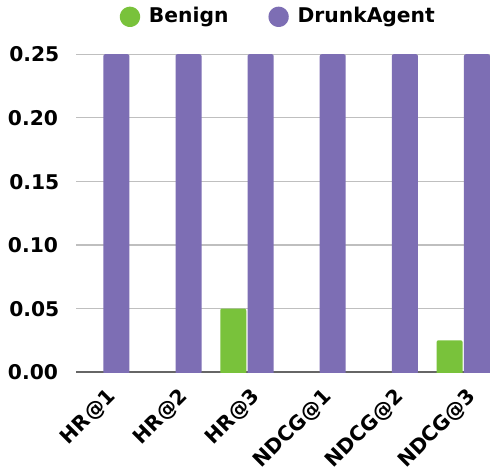}
    \caption{AgentSEQ}
    \label{yelp_agentseq}
\end{subfigure}
\vspace{-1.2em}
\caption{\textbf{Attack Transferability.} HRs and NDCGs of the Benign status and DrunkAgent against various black-box victim LLM-powered agent-based RSs on the real-world Yelp dataset.}
\label{fig:yelp_datasets}
\vspace{-1.4em}
\end{figure}

\subsubsection{Attack Generalizability across Domains} \label{cross_domain}

To further assess the generalizability of DrunkAgent, we conduct additional experiments on the Yelp dataset. Compared to the e-commerce-focused Amazon dataset, Yelp centers around user reviews of local businesses, including restaurants, cafes, bars, salons, repair shops, and other service-oriented establishments. It places a strong emphasis on experiential content, where user reviews play a central role in shaping recommendations. This domain is inherently different from traditional product recommendation: it often involves more subjective preferences and temporal dynamics (e.g., trends in dining or local services). These characteristics make Yelp a complementary and challenging testbed for evaluating the transferability and effectiveness of our method on different domains. Due to the expensive API calls, 772 interactions are randomly sample as a subset, including 61 users and 552 items. The sparsity of the subset is 97.71\%, which is more dense than the e-commerce ones. Despite these domain-specific differences, as can be seen from Fig.~\ref{fig:yelp_datasets}, DrunkAgent remains highly effective: it consistently promotes the target items on black-box LLM-powered recommender agents under memory corruptions, confirming the transferability of our attack and the effectiveness of memory perturbations across domains. This cross-domain success highlights the adaptability and generalizability of DrunkAgent, demonstrating that its impact is not limited to e-commerce settings but extends to real-world platforms driven by experiential content and local service discovery.

\subsubsection{Attack Interpretability} \label{adversarial_examples}

We focus on why adversarial attacks exhibit both transferability and stealthiness. We attribute the transferability of our adversarial texts to the use of a surrogate model that simulates the victim agent’s behaviors, such as memory updates and recommendations. By optimizing on this surrogate, the texts are equipped with transferable update masking strategies and are encapsulated with generalizable attack objectives (e.g., `top-level consideration', `a must-have', `prime choice') that can  effectively mislead similar models. Their stealthiness arises from maintaining semantic meaningfulness and coherence, which conceals manipulative intent and subtly disrupts prompt parsing.

\end{document}